\documentclass[amsfonts, amssymb, amsmath, reprint, showkeys,dvipsnames,longbibliography,superscriptaddress]{revtex4-2} 
\usepackage[english]{babel}
\usepackage[utf8]{inputenc}
\usepackage{amsthm}
\usepackage{mathtools}
\usepackage{physics}
\usepackage{xcolor}
\usepackage{graphicx}
\usepackage{adjustbox}
\usepackage{placeins}
\usepackage[T1]{fontenc}
\usepackage{lipsum}
\usepackage{csquotes}

\usepackage{xcolor}
\usepackage{hyperref}

\newcommand\myshade{85}
\colorlet{mylinkcolor}{Blue}
\colorlet{mycitecolor}{Blue}
\colorlet{myurlcolor}{Blue}

\hypersetup{
  linkcolor  = mylinkcolor!\myshade!black,
  citecolor  = mycitecolor!\myshade!black,
  urlcolor   = myurlcolor!\myshade!black,
  colorlinks = true,
}

\usepackage[version=4]{mhchem}
\usepackage{graphicx}
\usepackage{dcolumn}
\usepackage{bm}
\usepackage{bbding}
\usepackage{braket}
\usepackage{physics}
\usepackage{url}

\begin{document}
\title{Engineering the in-plane anomalous Hall effect in \ce{Cd3As2} thin films}

\author{Wangqian Miao}
\affiliation{Materials Department, University of California, Santa Barbara, California 93106-5050, USA}
\author{Binghao Guo}
\affiliation{Materials Department, University of California, Santa Barbara, California 93106-5050, USA}
\author{Susanne Stemmer}
\affiliation{Materials Department, University of California, Santa Barbara, California 93106-5050, USA}
\author{Xi Dai}
\email{daix@ust.hk}
\affiliation{Materials Department, University of California, Santa Barbara, California 93106-5050, USA}
\affiliation{Department of Physics, The Hong Kong University of 
Science and Technology, Clear Water Bay, Hong Kong, China}
\date{\today} 

\begin{abstract}
We predict two topological phase transitions for cadmium arsenide (\ce{Cd3As2}) thin films under in-plane magnetic field, taking advantage of a four-band $k\cdot p$ model and effective $g$ factors calculated from first principles. Film thickness, growth direction and in-plane Zeeman coupling strength can all serve as control parameters to drive these phase transitions. For (001) oriented \ce{Cd3As2} thin films, a two dimensional Weyl semimetal phase protected by $C_{2z}\mathcal{T}$ symmetry can be realized using an in-plane magnetic field, which has recently been reported in our companion paper. We then put forth two pathways to achieve in-plane anomalous Hall effects (IPAHE). By either introducing a trigonal warping term or altering the growth orientation, the emergent 
$C_{2z} \mathcal{T}$ symmetry can be broken. Consequently, in the clean limit and at low temperatures, quantized Hall plateaus induced by in-plane Zeeman fields become observable.
\end{abstract}


\maketitle

\section{Introduction}
Recent magnetotransport measurements \cite{guo_2d_weyl_2023, guo_2022, lygo_two-dimensional_2023} of cadmimum arsenide thin films (\ce{Cd3As2}) grown by molecular beam epitaxy (MBE) have revealed several attractive topological phases in two dimensions (2D), including a topological insulator phase and a Weyl semimetal phase. Notably, (001) grown \ce{Cd3As2} thin films under an in-plane Zeeman magnetic field offer a minimal model of a Weyl semimetal in 2D, where it hosts \textit{one single} pair of Weyl nodes carrying opposite chirality, separated in the momentum space, and protected by $C_{2z} \mathcal{T}$ symmetry.

Weyl semimetals are elusive in solid-state systems \cite{wan_weyl,ruan2016symmetry,armitage_weyl_2018}. To acquire this topological phase, either time reversal symmetry ($\mathcal{T}$) or space inversion symmetry ($\mathcal{P}$) must be broken. The first experimentally confirmed 3D Weyl semimetal \cite{Lv_TaAs_exp, Weng_TaAs_theory, xu2015discovery}, the non-centrosymmetric crystal \ce{TaAs} (which breaks $\mathcal{P}$), hosts 12 pairs of Weyl nodes near the Fermi surface, and displays exotic physical phenomena including the chiral anomaly and chiral zero sounds \cite{chiral_zero_sound_exp, chiral_zero_sound_song}. The large number of Weyl nodes near the Fermi level, however, has hindered detailed further investigations. A second route to a Weyl semimetal is to break time reversal symmetry ($\mathcal{T}$) starting with a parent Dirac semimetal state that has large $g$ factors. This approach has been discussed in Ref.~\cite{cano_chiral_2017, baidya_cd3as2_zeeman_prb}, and relies on an external magnetic field or magnetic dopants. The location of Weyl nodes may be controlled by the field strength, offering an experimental advantage. We point out here that the Zeeman effect and orbital effect are two major consequences of having an external magnetic field. It is not enough to only consider the Zeeman effect for $\textit{bulk}$ materials and ignore the orbital effect. Lastly, it was recently reported in Ref.~\cite{xiao_cd3as2_mag_doping} that magnetic doping remains challenging for bulk Dirac semimetals, such as \ce{Cd3As2}. 

It is natural to ask if we can find a system hosting a minimal number of Weyl nodes, and ideally in a low-dimensional material. Several materials are predicted to host a single pair of Weyl nodes \cite{you_two-dimensional_2019, MBT_weyl, single_weyl_Eucd2as2, zhao2022two, single_mag_weyl, weyl_nonmag, PhysRevB.109.075419}, but few have been experimentally confirmed. In 3D, while Weyl nodes are topologically stable against perturbations, external symmetry is required to protect the Weyl nodes in 2D materials. More specifically, as noted in Ref.~\cite{fang_new_2015, ahn_unconventional_2017}, a space-time inversion symmetry ($C_{2z}\mathcal{T}$) is the necessary component. The $C_{2z}\mathcal{T}$ symmetry enforces the Berry curvature to be zero everywhere in momentum space except at the band touching points that are the Weyl nodes. For example, in a \ce{Cd3As2} quantum well in the (001) direction, $C_{2z}$ symmetry is inherited from the $C_{4z}$ rotational symmetry  and the product symmetry $C_{2z}\mathcal{T}$ survives as long as the external magnetic field is directed in the film plane. The 2D Weyl semimetal phase is therefore predicted to appear \cite{burkov_phase_tran} when the quantum well subbands become inverted by the in-plane magnetic field. It is worth mentioning, that in the thin film region, the orbital effect of the magnetic field is greatly suppressed. The Zeeman effect, which is already significantly enhanced in systems with strong spin orbital coupling (SOC) \cite{gfactor_eng_prb_23}, becomes substantial and leads to the lifting of spin degeneracy. The spin-down subbands are pushed down by the magnetic field while the spin-up subbands are lifted up. As a result, \textit{one} pair of unobstructed Weyl nodes is created by the Zeeman field when the in-plane magnetic field is sufficiently strong and the zero-field band gap is small. On one side, these Weyl points can be killed under a small out of plane magnetic field. On the other side, one can imagine the $C_{2z} \mathcal{T}$ symmetry can be spontaneously broken, for example, when the thin film is grown along (112) direction. As a result, the Weyl points will be gapped out and the Berry curvatures are localized near their original positions, leading the bands below the Fermi level to acquire a finite Chern number $|\mathcal{C}| = 1$. This makes an in-plane anomalous Hall effect (IPAHE) or anomalous planar Hall effect (APHE) \cite{liu_-plane_2013,
sun_possible_2022, yao_inplane_qah, kurumaji_iphall_symm,ipahe_prl_21,ren_ipqah_prb_17, ipahe_jwang_prb_19,ding_ipahe_arxiv} feasible (For the terminology used here, see \footnote{We find that the term `in-plane anomalous Hall effect' is more appropriate for the phenomena described in this paper, as discussed in Ref.~\cite{kurumaji_iphall_symm}. However, it's worth noting that some papers refer to this as the `anomalous planar Hall effect' \cite{anomalous_planar_hall_prr, absence_aphe}.} ). These exotic quantum Hall signals emerge due to the non-trivial band topology tuned by the in-plane Zeeman field with appropriate symmetries, which is different from previous studies of so-called three dimensional quantum Hall effects \cite{3d_quantum_hall_lu, hinge_dirac_lu} in \ce{Cd3As2} thick slabs.

In this manuscript, we first present the semi-classical transport theory for thin films under in-plane mangetic field and the full $k \cdot p$ Hamiltonian for \ce{Cd3As2} with higher order corrections in Sec.~\ref{sec:theory}. The numerical studies of two Zeeman field induced topological phase transitions are shown in Sec.~\ref{sec:calculation}. We summarize our discovery and discuss possible experimental realization in Sec.~\ref{sec:summary}.

\section{General theory and model Hamiltonian \label{sec:theory}}

\subsection{Semi-classical transport theory}
We start by introducing the linear order response of the non-magnetic thin film system subjected to an in-plane magnetic field. This analysis will serve as a foundation for investigating the case of \ce{Cd3As2}. The semi-classical equation of motion for the Bloch electron wave packet can be formulated, taking into account the applied magnetic and electric fields \cite{rmp_niu} (we take $e=\hbar=1$), 
\begin{equation}
    \begin{aligned}
        \dot{\bm{r}} &= D^{-1} (\bm{v} + \mathbf{E} \cross \mathbf{\Omega}+ (\mathbf{\Omega}\cdot \bm{v})\mathbf{B}) \\
        \dot{\bm{k}} &= D^{-1} (-\mathbf{E} -  \bm{v} \cross \mathbf{B} - (\mathbf{E} \cdot \mathbf{B}) \mathbf{\Omega}),
    \end{aligned}
\end{equation}
where $\mathbf{\Omega}$ is the berry curvature, $D=1+ \mathbf{B} \cdot \mathbf{\Omega}$ is the volume factor, and $\bm{v}$ is the group velocity. The band index is neglected here. When the electrical field and magnetic field is co-planar, we assume $\mathbf{E} = (E, 0, 0)$ and $\mathbf{B} = B(\cos \theta, \sin \theta, 0)$. The linear order transverse conductivity $\sigma_{yx} = \sigma_{yx}^{(1)}+\sigma_{yx}^{(2)}$ can be solved under relaxation approximation \cite{chiral_anomaly_planar_hall_prl, weng_planr_hall, yao_planar_hall_prb_23}:
\begin{equation}
    \begin{aligned}
        \sigma_{yx}^{(1)} = &\tau \int [d\bm{k}] D^{-1} \left(-\frac{\partial f_{\text{eq}}}{\partial \varepsilon}\right) [v_x + B\cos \theta (\mathbf{\Omega}\cdot \bm{v})]\\
                            & \times [v_y+B\sin \theta(\mathbf{\Omega}\cdot \bm{v})] \\ 
        \sigma_{yx}^{(2)} = &\int[d \bm{k}] f_{\text{eq}} \mathbf{\Omega}_z ,
    \end{aligned}
\end{equation}
where $f_{\text{eq}}$ is the equilibrium Fermi-Dirac distribution
function with the energy dispersion $\varepsilon$ and high order contributions are ignored here. The first part $\sigma^{(1)}_{xy}$ contains traditional planar Hall response and Drude term, which scales linearly with relaxation time $\tau$. This contribution has been extensively studied in three dimensional topological Weyl semimetal and nodaline semimetal \cite{weng_planr_hall, yao_planar_hall_prb_23}. The second part $\sigma^{(2)}_{xy}$ is dissipationless and anti-symmetric which results from non-trivial Berry curvature. In experiments, these contributions to the transverse conductivity can be distinguished through taking measurements with $\mathbf{B}$ and $-\mathbf{B}$ and performing scaling analysis. In this manuscript, we mainly focus on how in-plane magnetic field will contribute to anomalous Hall conductivity $\sigma_{yx}^{(2)}$. One can imagine, in the two dimensional limit, the in-plane magnetic field will not produce Lorentz force and it only interacts with Bloch electrons through effective Zeeman coupling. This phenomenon only holds in the thin film region, where the film thickness is approximately 25 nm or less, and the magnetic length is comparable to the film thickness, typically for magnetic fields below 15 T. The quantized plateaus induced by the Zeeman effect in this system differ from the previously studied 3D quantum Hall effects in \ce{Cd3As2} thick slabs (with dimensions on the scale of 100 nm to 200 nm, as reported in Ref.~\cite{3d_quantum_hall_lu, hinge_dirac_lu}). In those studies, Landau levels were observed.

In this study, we present effective $g$ factors in the bulk $k \cdot p$ model using first principle wave functions following Ref.~\cite{song_chapter_2020,sun_topological_2020}. Consequently, we can simultaneously capture the in-plane magnetic field-induced corrections for band dispersion, group velocity and Berry curvature, which were previously treated as a perturbative $O(B)$ correction in Ref.~\cite{rmp_niu, yang_gao_prl_bcp,non_linear_inplane_mag_prb, IPHE_orbital_prl}.

\subsection{Low energy effective model}
We first revisit the low energy effective model for bulk \ce{Cd3As2}, which belongs to tetragonal crystal system. The effective orbitals near the Fermi surface can be represented as $\ket{\frac{1}{2}, \pm \frac{1}{2}}$ and $\ket{\frac{3}{2}, \pm \frac{3}{2}}$ and the symmetry allowed low energy \cite{wang_three-dimensional_2013,cano_chiral_2017, baidya_cd3as2_zeeman_prb} $ k\cdot p$ Hamiltonian then reads: 
\begin{equation}
    \mathcal{H}_0(\bm{k})= \epsilon(\bm{k})+
    \begin{pmatrix}
        \mathcal{M}(\bm{k}) &Dk_{-} &\mathcal{A}^{*}(\bm{k}) &\mathcal{B}^*(\bm{k})\\
        Dk_{+} &\mathcal{M}(\bm{k}) &\mathcal{B}(\bm{k}) &-\mathcal{A}(\bm{k}) \\
        \mathcal{A}(\bm{k})&\mathcal{B}^*(\bm{k}) &-\mathcal{M}(\bm{k}) &0 \\
        \mathcal{B}(\bm{k}) &-\mathcal{A}^{*}(\bm{k}) &0 &-\mathcal{M}(\bm{k}) 
    \end{pmatrix},
    \label{eq:kp}
\end{equation}

where $\epsilon(\bm{k}) = C_0 + C_1 k_z^2 + C_2 (k_x^2+k_y^2 )$,  $\mathcal{M}(\bm{k}) = M_0+ M_1 k_z^2+ M_2 (k_x^2+k_y^2 ), \mathcal{A}(\bm{k}) = Ak_{-} + B_{1} (k_x^3-ik_y^3)+ i B_2 k_{+}k_x k_y  - B_5 k_{-}k_z^2,  \mathcal{B}(\bm{k}) = B_3(k_x^2-k_y^2)k_z+iB_4 k_x k_y k_z, k_{\pm} = k_x \pm i k_y$. $D k_{\pm}$ term breaks the inversion symmetry and cubic terms characterize the $C_{4z}$ rotational symmetry. The $x,y, z$ axis of the Hamiltonian are defined along (100), (010), (001) directions. Up to the second order, this Hamiltonian hosts two Dirac points located at $(0, 0, \pm\sqrt{-M_0/M_1})$. In bulk, single crystals of \ce{Cd3As2}, structures with and without an inversion center have been observed, as reported in \cite{ali_crystal_2014,Sankar2015-je}, which demonstrate that growth and processing conditions determine the specific room-temperature crystal structure. In this manuscript, we mainly focus on the one with inversion symmetry ($D k_{\pm}=0$), which is based on experiments that show the preservation of inversion centers in high-quality MBE-grown \ce{Cd3As2} thin films on III-V substrates. This point has been carefully examined using convergent beam electron diffraction technique in Ref.~\cite{guo_2022}. Note that, the $k \cdot p$ Hamiltonian possesses a $C_{\infty}$ rotational symmetry in (001) direction without cubic terms. Furthermore, the Zeeman coupling term for the bulk $k\cdot p$ model can be constructed \cite{liu_model_2010}:
\begin{equation}
    \mathcal{H}_{\text{Zeeman}} = 
\mu_{\mathrm{B}} 
    \begin{pmatrix}
        g_{1z} B_z &g_{1p} B_- &0 &0 \\
        g_{1p} B_{+} &-g_{1z}B_z &0 &0\\
        0 &0 &g_{2z}B_z & g_{2p}B_{+} \\
        0 &0 &g_{2p}B_{-} &-g_{2z} B_z
    \end{pmatrix}
    \label{eq:zeeman}
\end{equation}

where the magnetic field is chosen as $\mathbf{B}=(B_x, B_y, B_z)$ and $B_{\pm} = B_x \pm i B_y$. We adopt the following parameters for the $k \cdot p$ model, $C_0$ = -0.0145 eV, $C_1$ = 10.59 eV \AA, $C_2$=
11.5 eV \AA$^2$, $M_0$ = -0.0205 eV, $M_1$ = 18.77 eV \AA$^2$, $M_2$ =
13.5 eV \AA$^2$, $A$ = 0.889 eV \AA  \cite{cano_chiral_2017}. $B_1 = 0.0153$ eV \AA$^3$, $B_2 = -0.0476$ eV \AA$^3$, $B_3 = 0.0669$ eV \AA$^3$, $B_4 = -0.0366$ eV \AA$^3$, $B_5 = 0.0668$ eV \AA$^3$, $g_{1p} = 11.6211$, $g_{2p} = 0.5876, g_{1z} = 10.0316, g_{2z} = -4.3048$ are new parameters calculated using first-principles wave-functions following the procedures described in Ref.~\cite{song_chapter_2020, sun_topological_2020}. To acquire the band dispersion of \ce{Cd3As2} thin film grown in a general direction, denoted as $z'$, we rotate $z$ axis to $z'$ axis using the following matrix:
\begin{equation}
    \begin{pmatrix}
        k_x' \\
        k_y' \\
        k_z'
    \end{pmatrix} = 
    \begin{pmatrix}
        &\cos \theta &-\sin \theta &0 \\
        &\sin \theta \cos \phi &\cos \theta \cos \phi &-\sin \phi \\
        &\sin \theta \sin \phi &\cos \theta \sin \phi &\cos \phi
    \end{pmatrix}
    \begin{pmatrix}
        k_x\\
        k_y\\
        k_z
    \end{pmatrix}.
\end{equation}
we set $(\theta, \phi) = (0, 0), (\pi/4, \arctan(\sqrt{2}/2))$ for \ce{Cd3As2} thin films grown along the (001) and (112) directions, respectively, which are available through MBE. 

\subsection{Critical thickness for Cd$_3$As$_2$ thin films}
\begin{figure}
\includegraphics[width=0.48\textwidth]{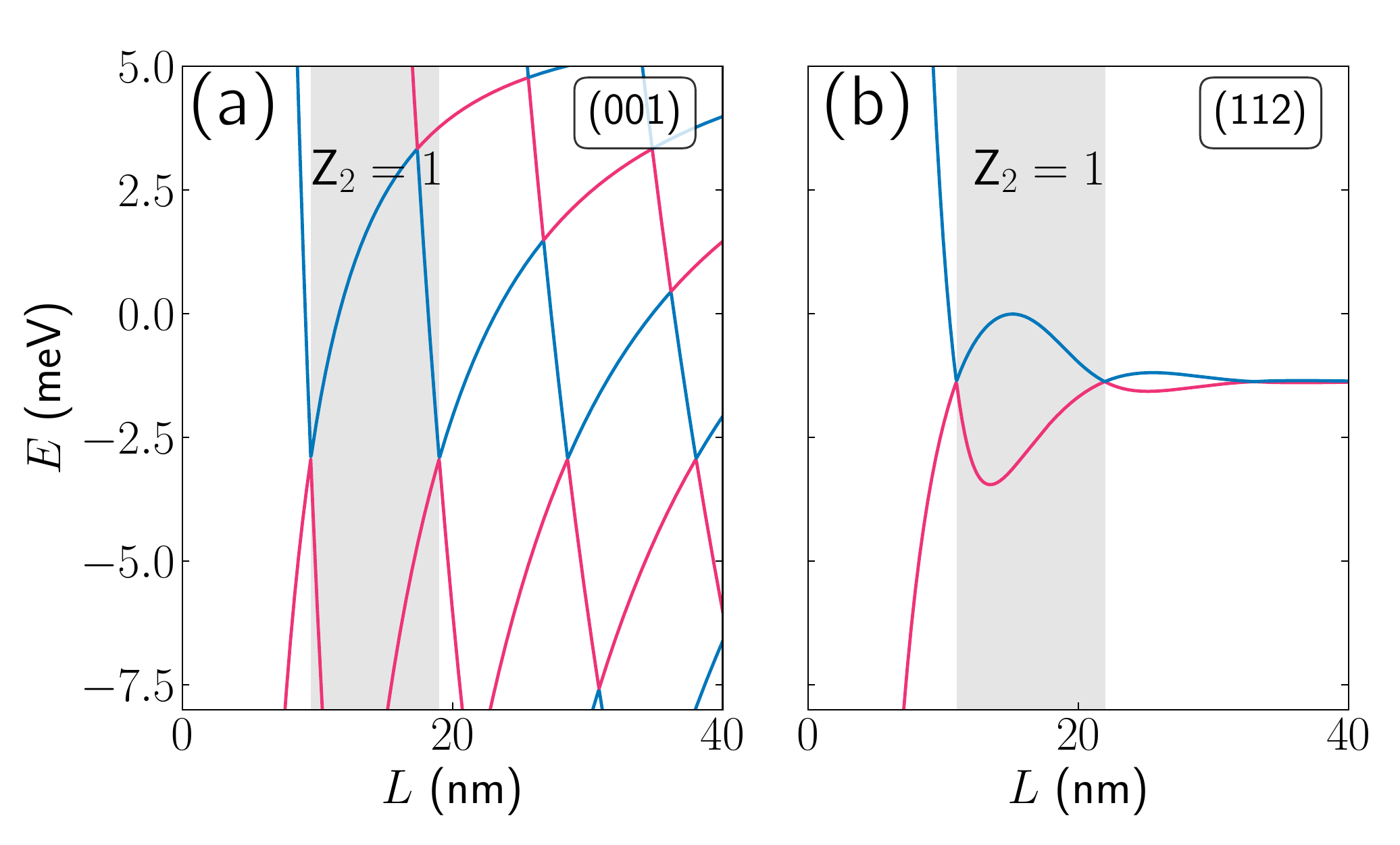}
    \caption{\label{fig:fig1} Energy levels at the $\Gamma$ point near the Fermi surface  versus the thin films thickness for \textbf{a.} (001) thin films, \textbf{b.} (112) thin films. Here, we take $D=0$ and ignore cubic terms in Eq.~(\ref{eq:kp}). The 2D-TI phase appears in a oscillatory fashion as the function of the thickness of the thin film. We labeled the first 2D-TI ($\mathrm{Z}_2=1$) region in the shadowed area, which has the best chance to be realized in experiments.}
\end{figure}

\begin{figure*}
\includegraphics[width=0.9\textwidth]{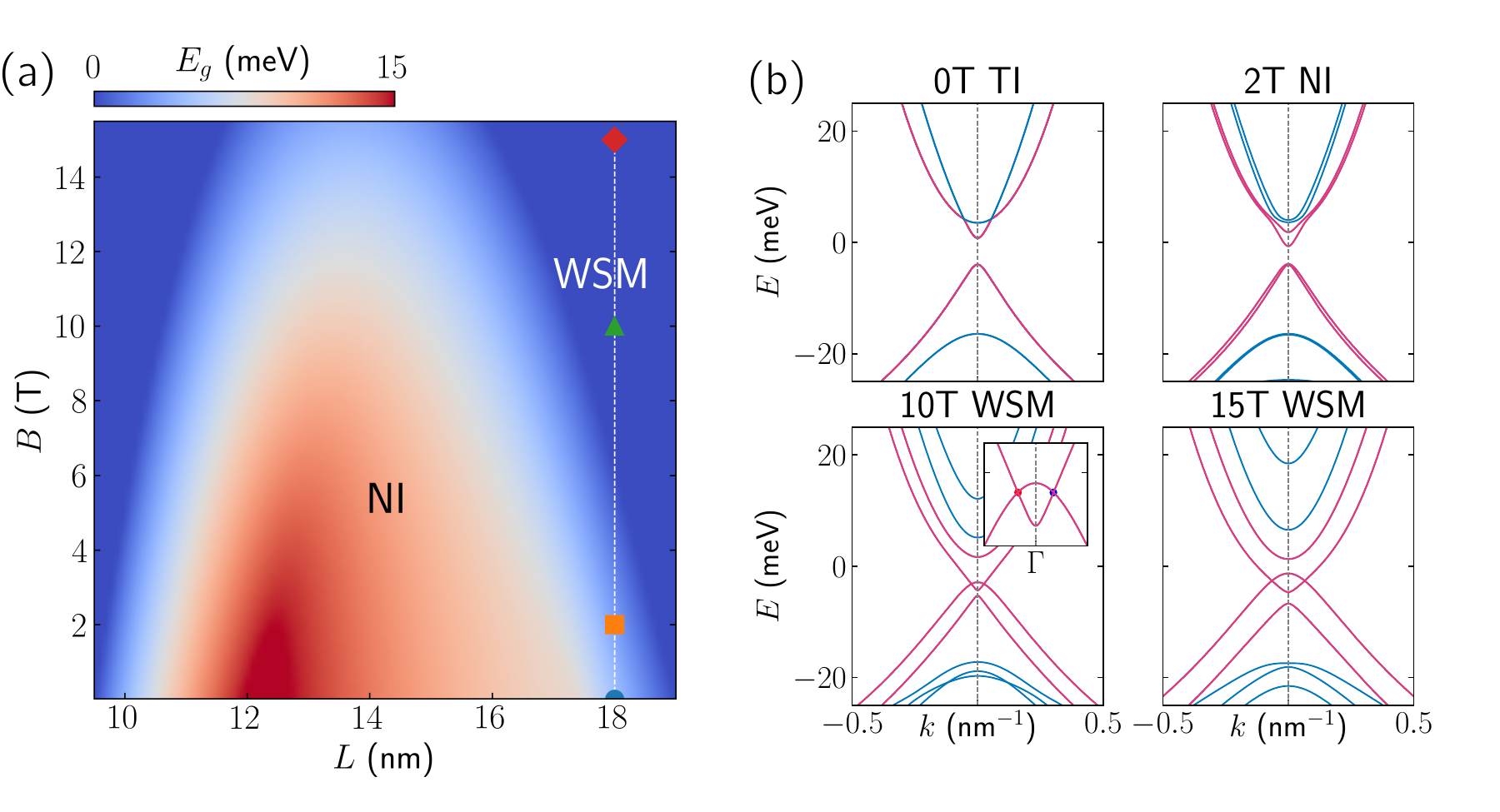}
    \caption{\label{fig:001-film} (a) Topological phase diagram for (001) grown \ce{Cd3As2} thin films under in-plane magnetic field $B$. The colorbar labels the calculated band gap of the thin films at the $\Gamma$ point. We focus on the 9 nm to 19 nm-thick thin films which are in the 2D-TI phase when there is no magnetic field. The normal insulator (NI) phase and  Weyl semimetal (WSM) phase are labeled in different colored regions. We show the most possible thickness to realize WSM phase is close to the critical point (19 nm). For an 18 nm thin film, we take a line cut on the phase diagram to illustrate different phases. The half-circle denotes the 2D-TI phase (0 T), the square denotes the normal insulator phase (2 T), the triangular is the WSM phase (10 T) and the diamond is the WSM phase (15 T). (b) Band structures calculated along the direction perpendicular to the magnetic field for 18 nm (001) grown thin films under in-plane magnetic field. Zeeman effect and orbital effect are all considered. Four central bands close to the Fermi level are depicted in red while other bands are in blue. When $B = 0$ T, the system is in a 2D-TI phase, all bands are doubly degenerate. When $B = 2$ T, the system is a normal insulator. When $B= 10, 15$ T, the system is in a WSM phase. The inset shows the band dispersion of the 2D WSM, but within a lower energy range. Two Weyl nodes are separated in the momentum space and a stronger in-plane magnetic field can push them away from each other.}
\end{figure*}

We emphasize that in order to achieve the in-plane Zeeman field-induced topological transition, two conditions must be satisfied: the film thickness should be sufficiently thin, and the band gap should be small. By meeting these criteria, the band inversion can occur through the application of a moderate magnetic field. Therefore, it becomes crucial to determine the critical thickness of the thin film for different growth orientations. It was first proposed in Ref.~\cite{chaoxing_2DTI_crossover, lu_ti_2d, finite_size} that the electron band and hole band will oscillate with the film thickness and two dimensional topological insulator (2D-TI) phase can be obtained as a result of the quantum confinement effect. This is the case for Dirac semimetal \ce{Cd3As2} \cite{wang_three-dimensional_2013}. We start by presenting the sub-band structures of \ce{Cd3As2} thin films. To simulate the corresponding quantum well (QW) structure in experiments, the continuous limit that $k_z' = -i \partial_{z'}$ and open boundary condition along $z'$ directions are applied. The Hamiltonian $\mathcal{H} (\bm{k}') = \mathcal{H}(k_x', k_y', -i\partial_{z'})$ can then be expanded using a plane-wave basis

\begin{equation}
        \psi_n (z') = \sqrt{\frac{2}{L}} \sin \left[ \frac{n\pi}{L} \left(z'+\frac{L}{2}\right) \right], -L/2<z'<L/2,
        \label{eq:basis}
\end{equation}
where $L$ is the thickness of the thin film and $n$ is taken up to 50. As shown in Fig.~\Ref{fig:fig1}, the electron band and hole band oscillate, which allows us to obtain the 2D TI phase by tuning the thin film thickness. We first focus on the thin films grown along (001) direction. By direct inspection of Eq.~(\Ref{eq:kp}) (set $D=0$ and ignore cubic terms), we can easily solve for the energy levels at $\Gamma$ point with $E_n= C_0+M_0+(C_1+M_1)(n\pi/L)^2$, $H_n=C_0-M_0+(C_1-M_1)(n\pi/L)^2$. The critical thickness $L_{cn}$ should satisfy $E_n = H_n$. Then we get $L_{cn} = n \pi /\sqrt{-M_0/M_1}$ and the same analysis can be performed for (112) grown thin films. For (001) thin films, $L_{c1}=10$nm, $L_{c2}=19$nm, for (112) thin films,  $L_{c1} = 11$ nm, $L_{c2} = 21$ nm. We labeled the first 2D-TI region in Fig.~\Ref{fig:fig1} and the numerical result is consistent with recent transport experiment \cite{lygo_two-dimensional_2023}. Among different thicknesses, a thin film grown around 20 nm has the highest probability of realizing Zeeman field induced effect. In the upcoming discussion, we will delve into the details of this phenomenon.

\section{Zeeman-field induced physical effects \label{sec:calculation}}

In this section, we demonstrate that \ce{Cd3As2} thin films can manifest rich topological phase diagrams under the influence of an in-plane magnetic field. The Zeeman effect serves to lift the spin degeneracy of the sub-bands, leading to their inversion when the in-plane magnetic field is sufficiently strong. For thin films grown along the (001) direction, a single pair of Weyl points can be stabilized by $C_{2z} \mathcal{T}$ symmetry. Once this symmetry is broken, the emergence of the in-plane quantum anomalous hall effect can be anticipated. Conversely, for thin films grown along the (112) direction and near a critical thickness, a topological gap remains robust when an in-plane Zeeman field is introduced.


\subsection{Two dimensional Weyl semimetal phase}

We start with a general analysis for (001) grown \ce{Cd3As2} thin films under in-plane magnetic field. We can expect the orbital effect (which results in Landau Levels) and Zeeman effect to co-exist when the field is switched on. We first estimate the magnetic length $\ell_{\mathrm{B}} =\sqrt{\hbar/eB}= 25.6 /\sqrt{B[\text{T}]} [\text{nm}]$ is around 10 nm for 10 T which is comparable with the thin film thickness. The cyclotron motion of the electrons along (001) direction is greatly suppressed by the quantum confinement effect. On the other side, the Zeeman effect leads to the spin splitting and the energy levels at $\Gamma$ point can be modified by a magnitude of $g\mu_{\mathrm{B}}B$. It is noteworthy that the energy scale of the Zeeman splitting is typically small in comparison to the large band gap found in conventional semiconductors. However, the band gap in these topological thin films is readily tunable, rendering the Zeeman effect particularly significant. As a result, the topological gap of the 2D-TI will be inverted. One pair of Weyl nodes appear at the band crossing point and they are protected by the new emergent $C_{2z}\mathcal{T}$ symmetry. This new symmetry guarantees the Berry curvature to be zero everywhere in the momentum space except the Weyl nodes and the Berry phase integrated along the path enclosing the Weyl nodes must be $\pm \pi$.

We confirmed above analysis using numerical calculations. We first assume $\mathbf{B} = (B_x, B_y,0), B_x = B \cos \varphi, B_y = B \sin \varphi$ and set $D=0, B_{1-5}=0$ for simplicity, the full Hamiltonian incorporating in-plane magnetic field should be:
\begin{equation}
    \mathcal{H}_{\mathrm{B}} = \mathcal{H}_0 \left(k_x'+ \frac{z'}{\ell_{\mathrm{B}_y}^2}, k_y'-\frac{z'}{\ell_{\mathrm{B}_x}^2}, -i\partial_{z'} \right) + \mathcal{H}_{\mathrm{Zeeman}},
\end{equation}
where Landau gauge and Peierls substitution are applied to fully consider orbital effect. Then the magnetic Hamiltonian $\mathcal{H}_{\mathrm{B}}$ can be expanded under the plane-wave basis $\psi_n(z')$ defined in Eq.~(\Ref{eq:basis}). The computed phase diagram illustrating the band gap as a function of the in-plane magnetic field 
$B$ and the film thickness $L$ is presented in Fig.\ref{fig:001-film}-(a). Notably, when the film thickness approaches the critical point $L_c$, even a moderate magnetic field can induce a phase transition from a normal insulator (NI) to a Weyl semimetal (WSM). For instance, consider an 18 nm thin film, as shown by the line cut in Fig.\ref{fig:001-film}-(a) and the corresponding band structures in Fig.\ref{fig:001-film}-(b). In the 2D-TI phase without an applied magnetic field, all sub-bands are doubly degenerate. With a sufficiently strong in-plane magnetic field (around 10 T), a pair of Weyl nodes emerges in the direction perpendicular to the magnetic field. We also explore the role of orbital effects in our calculations. Both Zeeman and orbital effects are accounted for in the sub-band structures depicted in Fig.\ref{fig:001-film}-(b). Our findings indicate that the orbital effect is suppressed in this thin film due to quantum confinement, meaning no emergence of Landau levels, see more numerical results in Supplementary Material (SM). Nonetheless, the orbital effect acts to spatially separate the Weyl nodes in momentum space. This 2D Weyl semi-metal phase is stable against inversion symmetry broken term $D k_{\pm}$ and cubic terms obeying $C_{4z}$ rotational symmetry in Eq.(\ref{eq:kp}), because $C_{2} \mathcal{T}$ symmetry survives after considering these perturbation \cite{fang_new_2015}, more numerical results considering cubic terms correction and inversion symmetry breaking term are shown in SM. 

\begin{figure}
\includegraphics[width=0.48\textwidth]{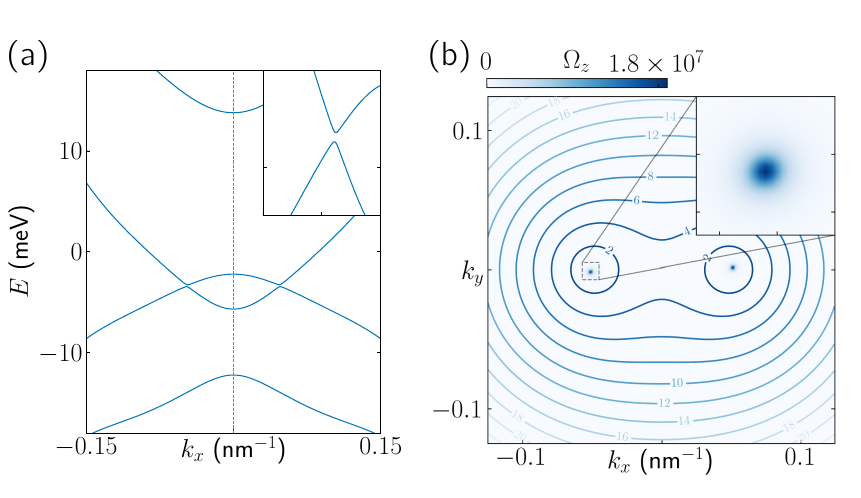}
    \caption{\label{fig:qah-weyl}
    (a) The band structure for the 2D Weyl semi-metal phase ($B_y = 15$ T) after integrating the trigonal warping term. ($R_1, R_2, R_3, R_4 = 100$ eV \AA$^{3}$) A small non-trivial gap is opened, as shown in the inset. (b) Calculated Berry curvature $\Omega_z(\bm{k})$ for all bands below Fermi level in the momentum space. The trigonal warping results in the singular behaviour of the Berry curvature. As depicted in the inset, the berry curvature concentrates near the original Weyl points. The contours for the band gap is also depicted in units of meV.}
\end{figure}

\subsection{In-plane anomalous Hall effect}
\begin{figure*}
\includegraphics[width=0.9\textwidth]{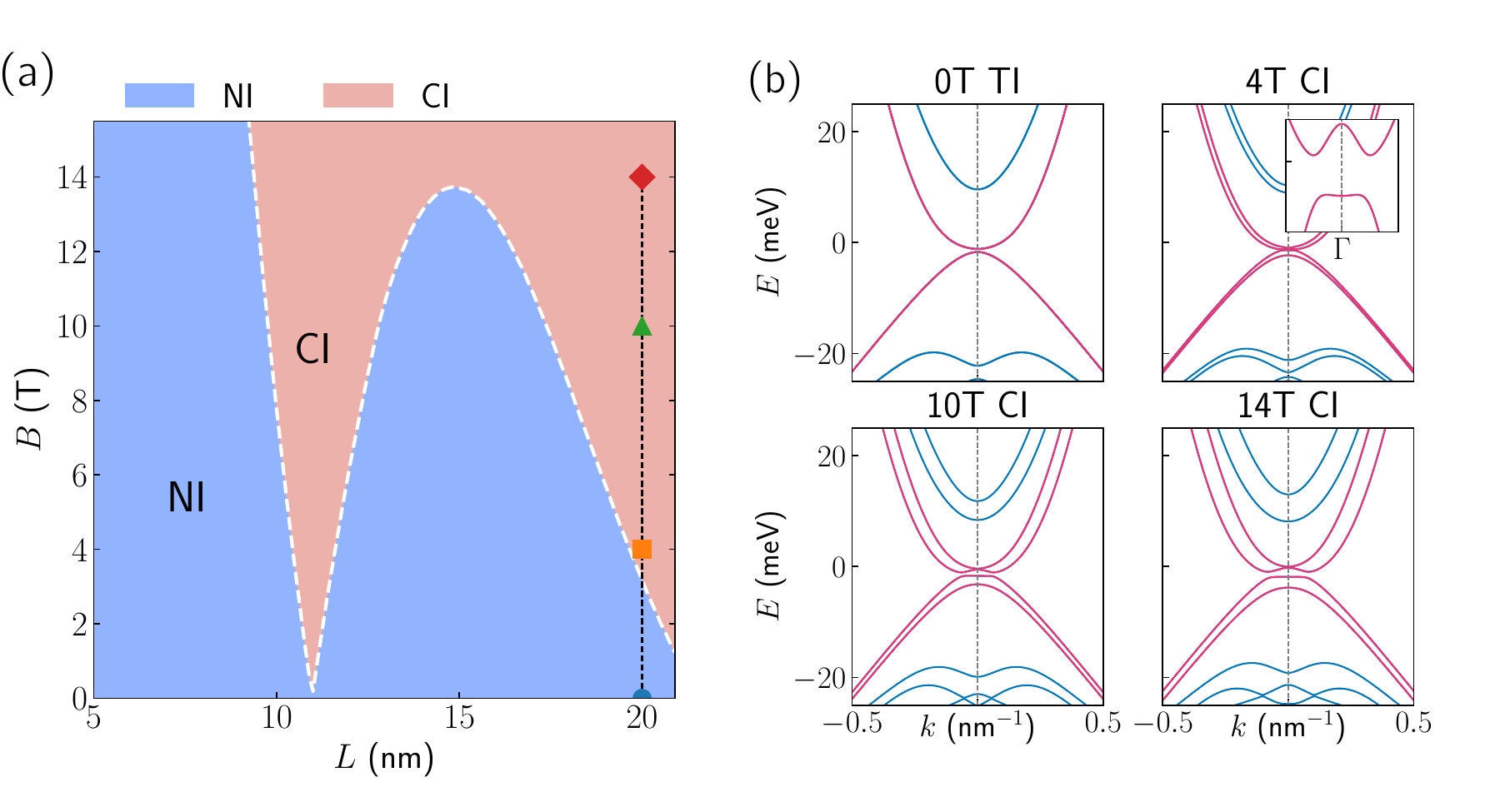}
    \caption{\label{fig:112-film} (a) Topological phase diagram for (112) \ce{Cd3As2} thin films under in-plane magnetic field. The normal insulator (NI) phase and  Chern insulator (CI) phase are labeled in different colored regions. The white dashed curve represents the minimal in-plane magnetic field to realize CI phase at each thickness. We show the most possible thickness to realize CI phase is close to the critical point (21 nm). For a 20 nm thin film, we take a line cut on the phase diagram to illustrate the different phases. The circle denotes the 2D-TI phase (0 T), the square (4 T), the triangular (10 T) and the diamond (14 T) are all in the CI phase but with increasing topological gaps. (b) Calculated band structures along $k_y'$ for a 20 nm (112) grown \ce{Cd3As2} thin film under different in-plane magnetic fields. Four bands close to the Fermi surface are labeled in red while other remote bands are in blue. When $B = 0$ T, it is in a 2D-TI phase, all bands are doubly degenerate. When $B \geq 14$ T, the system is in a CI phase. The topological gap persists and is enhanced by the in-plane magnetic field.}
\end{figure*}
The conventional quantum Hall effect arises due to the formation of Landau levels and the breaking of time-reversal symmetry in the presence of an external magnetic field. A quantized Hall conductance can also manifest in a specific class of materials, even in the absence of an applied magnetic field. This occurs when time-reversal symmetry is spontaneously broken, either by magnetic dopants \cite{chang2013experimental}, intrinsic magnetic order \cite{deng2020quantum}, or moir\'e potentials \cite{sharpe2019emergent,serlin2020intrinsic,chen2020tunable,li2021quantum}. This phenomenon is termed the quantum anomalous Hall (QAH) effect \cite{chang_liu_rmp}. As discussed in the previous section, the orbital effects of an in-plane magnetic field are suppressed in topological thin films with strong spin-orbit coupling (SOC). Concurrently, an enhanced Zeeman effect can induce the inversion of a pair of electron and hole band when the field is sufficiently strong. As a result, one could anticipate the emergence of a Chern insulator phase originating from the Zeeman effect upon the breaking of certain symmetries \cite{liu_-plane_2013, sun_possible_2022, yao_inplane_qah, kurumaji_iphall_symm}, an effect which is likewise \textit{anomalous}. 

The key principle for achieving a CI with an in-plane magnetic field in a 2D WSM we elaborate is to break the $C_{2z}\mathcal{T}$ symmetry. Upon the inclusion of symmetry-breaking terms, specifically trigonal warping \cite{liang_hex}. For the case of (001) grown \ce{Cd3As2}, it can be written as:
\begin{widetext}
\begin{equation}
    \label{eq:trigonal}
    \mathcal{H}_{\text{tri}} (\bm{k}) = 
    \begin{pmatrix}
        0 &0 &(R_3+iR_4)k_{+}k_{\parallel} &(R_1+iR_2)k_{+}k_{\parallel} \\
         0 &0 &(R_1-iR_2)k_{-}k_{\parallel} &(-R_3+iR_4)k_{-}k_{\parallel}\\
          (R_3-iR_4)k_{-}k_{\parallel} &(R_1+iR_2)k_{+}k_{\parallel} &0 &0 \\
         (R_1-iR_2)k_{-}k_{\parallel} &(-R_3-iR_4)k_{+}k_{\parallel} &0 &0 \\   
    \end{pmatrix},
\end{equation}    
\end{widetext}
where $k_\parallel = k_x^2+k_y^2$ and $R_1,R_2,R_3,R_4$ are real numbers. After integrating the trigonal warping in the minimal 2D WSM model, a non-trivial topological gap appears, as shown in Fig.~\ref{fig:qah-weyl}-(a). The Berry curvature $\Omega_z(\bm{k})$ shows a localized distribution near the initial positions of the Weyl nodes, as illustrated in Fig.\ref{fig:qah-weyl}-(b). The corresponding integral of $\Omega_z(\bm{k})$ becomes quantized to $2\pi$. From an experimental perspective, this kind of warping effect can be introduced in (001) grown thin films through introducing $C_{3}$ symmetric moir\'e potential \cite{moire_surf_fu,cano_moire_surface,moire_pont_cano,wan_moire_strain, moire-dirac_lin_prr_2023, miao_truncated} through gate engineering \cite{sun2023signature} or proximity effect \cite{Kim2023_thbn}, see a brief discussion in SM.

Then we focus on a realistic scenario to realize Zeeman field induced IPAHE in \ce{Cd3As2}. We further observe that thin films of \ce{Cd3As2} grown along the (112) direction, which corresponds to the (111)-axis in an approximate cubic unit cell, offer an exceptional platform. It is noteworthy that unlike the (001)-grown films, those grown along the (112) direction possess only an approximate \(C_3\) rotational symmetry. This naturally breaks the \(C_{2z}\mathcal{T}\) symmetry when an in-plane magnetic field is applied. To account for this type of symmetry breaking, we take the cubic terms in Eq.~(\ref{eq:kp}) into consideration. The topological phase diagram for these (112)-grown \ce{Cd3As2} thin films under an in-plane magnetic field is presented in Fig.~\Ref{fig:112-film}-(a). Note that, in the unrotated coordinate, $B_x+B_y+2B_z=0$ should be satisfied. If $\varphi$ captures the angle between the magnetic field and $k_x'$ axis, we get $B_x = (\sqrt{2}/6+\sqrt{2}/3) B\cos\varphi+1/\sqrt{3}B\sin\varphi,B_y = -(\sqrt{2}/6+\sqrt{2}/3) B\cos\varphi+1/\sqrt{3}B\sin\varphi, B_z = -1/\sqrt{3} B \sin \varphi $. The white dashed line in the diagram indicates the minimum magnetic field magnitude required to drive the system into the CI phase. The most favorable region for this transition is close to the critical thickness where the band gap is nearly zero. To illustrate the Zeeman effect-induced phase transition, we take a 20 nm thin film and set $\varphi=\pi/15$ as an example. In the absence of a magnetic field, the thin film behaves as a 2D TI, as shown in Fig.~\Ref{fig:112-film}-(b). Application of a small in-plane magnetic field leads to band inversion. Unlike the 2D WSM phase observed in (001)-grown thin films, the avoided band crossing is highlighted in Fig.~\Ref{fig:112-film}-(b). Increasing the magnitude of the in-plane magnetic field enlarges the topological gap. When \( B \geq 14 \) T, a sizable gap opens, facilitating the quantized IPAHE. 

\begin{figure*}
\includegraphics[width=0.9\textwidth]{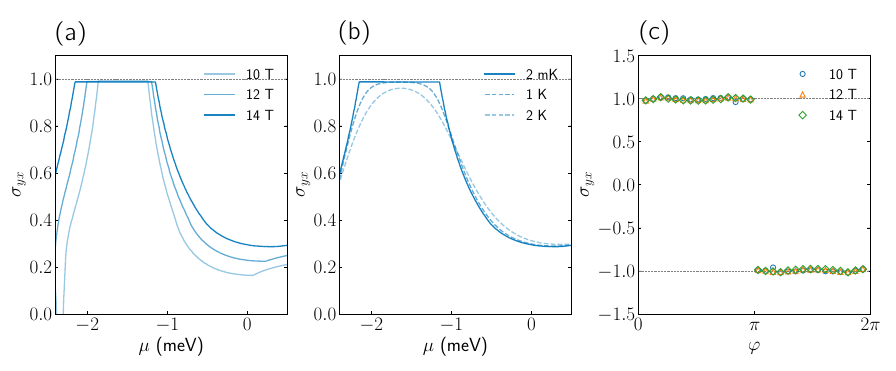}
    \caption{\label{fig:fig6_quantized}
    (a) Calculated Hall conductance $|\sigma_{yx}|$ versus the chemical potential at 2 mK for a 20 nm (112) grown thin film, with different magnitudes of applied in-plane magnetic fields ($\varphi=\pi/15$). (b) Calculated Hall conductance at different temperatures versus the chemical potential for a 20 nm (112) grown thin film applied with a 14 T in-plane magnetic field. The quantized plateaus are stable at low temperatures. (c) Quantized anomalous Hall conductance (chemical potential is set in the gap) versus the angle between the in-plane magnetic filed $B$ and $k_x'$. $B$ is set at 10/12/14 T. It is clear that $\sigma_{yx}(B) = -\sigma_{yx}(-B)$.}
\end{figure*}

We now investigate the impact of in-plane magnetic field strength and temperature on the quantized Hall conductance plateau in a 20 nm thin film using Kubo formula. In numerics, we preserve 32 subbands, use a $1500 \times 1500$ $k-$mesh and truncate the momentum space in a square region ($-0.8$ nm$^{-1}$ $\leq k_x', k_y' \leq$ 0.8 nm$^{-1}$). The calculated plateau for Hall conductance is close to quantized. As depicted in Fig.~\Ref{fig:fig6_quantized}(a)-(b), the width of the plateau increases with the application of a stronger magnetic field $B \geq 10$ T and is stabilized at an ultra-low temperature \( T \leq 1 \) K. We here emphasize that the IPAHE is distinct from the conventional planar Hall effect. Specifically, the Hall conductance for IPAHE is an odd function of the in-plane magnetic field, as opposed to an even function in the conventional case \cite{anomalous_planar_hall_prr, kurumaji_iphall_symm, absence_aphe}. This is also confirmed in our numerics that the sign of the Hall conductance will be changed by varying the direction (varying $\varphi$) of the in-plane magnetic field, as shown in Fig.~\ref{fig:fig6_quantized}. 
We further notice that a small deviation from inversion symmetry may cause noteworthy Berry curvature dipole distribution (We present the profile for the distribution of Berry curvature and corresponding dipole in SM ) which hints non-linear planar Hall effect \cite{non_linear_inplane_mag_prb,intrinsic_nonlinear_planar_hall_prl}  can co-exist with the quantized IPAHE. The breaking of inversion symmetry can be achieved through using specific growing methods or applying a vertical electrical field \cite{ gfactor_eng_prb_23}. This intriguing phenomenon will be the subject of future studies.

\section{Conclusions \label{sec:summary}}
In this paper, we conduct a comprehensive investigation of Zeeman-field induced topological phase transitions in \ce{Cd3As2} thin films. Unlike previous studies \cite{cano_chiral_2017,baidya_cd3as2_zeeman_prb} that mainly focus on the Zeeman effect in \textit{bulk} materials, we highlight the critical role of Zeeman splitting in thin films, especially when coupled with quantum confinement effects. Our analysis reveals that in thin films grown along the (001) direction, the orbital effects of an in-plane magnetic field \(B \leq 15\) T are negligible, and the Zeeman effect becomes the dominant factor driving topological phase transitions. For (001)-oriented films, we predict the emergence of a 2D Weyl semimetal (WSM) phase, which is stabilized by \(C_{2z}\mathcal{T}\) symmetry. Furthermore, the positions of the Weyl nodes can be manipulated through the direction and magnitude of the applied in-plane magnetic field. Once the \(C_{2z}\mathcal{T}\) symmetry is broken due to trigonal warping, it will give rise to Chern bands. On a parallel note, for thin films grown along the (112) direction where $C_{2z} \mathcal{T}$ is obstructed from geometry, we foresee the possibility of observing an IPAHE at low temperatures and in the clean limit, under a strong magnetic field. Our approach for achieving these phase transitions is generally applicable to MBE-grown topological thin films characterized by strong spin-orbit coupling, large effective \(g\)-factors, and appropriate symmetries \cite{yao_inplane_qah, kurumaji_iphall_symm}.

\begin{acknowledgments}
W. Miao thanks S. Zhang, S. Sun and H. Shi for valuable discussions and acknowledges travel award granted from UCSB NSF Quantum Foundry funded via DMR-1906325. Part of this work made use of the computational facilities supported by the U.S. NSF (CNS-1725797) and administered by the Center for Scientific Computing.  B. Guo acknowledges support from the Graduate Research Fellowship Program of the U.S. NSF (Grant No. 2139319), and the UCSB Quantum Foundry, which is funded via the Q-AMASE-i program of the U.S. NSF (Grant No. DMR-1906325).
\end{acknowledgments}

\bibliography{reference}

\begin{thebibliography}{71}%
\makeatletter
\providecommand \@ifxundefined [1]{%
 \@ifx{#1\undefined}
}%
\providecommand \@ifnum [1]{%
 \ifnum #1\expandafter \@firstoftwo
 \else \expandafter \@secondoftwo
 \fi
}%
\providecommand \@ifx [1]{%
 \ifx #1\expandafter \@firstoftwo
 \else \expandafter \@secondoftwo
 \fi
}%
\providecommand \natexlab [1]{#1}%
\providecommand \enquote  [1]{``#1''}%
\providecommand \bibnamefont  [1]{#1}%
\providecommand \bibfnamefont [1]{#1}%
\providecommand \citenamefont [1]{#1}%
\providecommand \href@noop [0]{\@secondoftwo}%
\providecommand \href [0]{\begingroup \@sanitize@url \@href}%
\providecommand \@href[1]{\@@startlink{#1}\@@href}%
\providecommand \@@href[1]{\endgroup#1\@@endlink}%
\providecommand \@sanitize@url [0]{\catcode `\\12\catcode `\$12\catcode `\&12\catcode `\#12\catcode `\^12\catcode `\_12\catcode `\%12\relax}%
\providecommand \@@startlink[1]{}%
\providecommand \@@endlink[0]{}%
\providecommand \url  [0]{\begingroup\@sanitize@url \@url }%
\providecommand \@url [1]{\endgroup\@href {#1}{\urlprefix }}%
\providecommand \urlprefix  [0]{URL }%
\providecommand \Eprint [0]{\href }%
\providecommand \doibase [0]{https://doi.org/}%
\providecommand \selectlanguage [0]{\@gobble}%
\providecommand \bibinfo  [0]{\@secondoftwo}%
\providecommand \bibfield  [0]{\@secondoftwo}%
\providecommand \translation [1]{[#1]}%
\providecommand \BibitemOpen [0]{}%
\providecommand \bibitemStop [0]{}%
\providecommand \bibitemNoStop [0]{.\EOS\space}%
\providecommand \EOS [0]{\spacefactor3000\relax}%
\providecommand \BibitemShut  [1]{\csname bibitem#1\endcsname}%
\let\auto@bib@innerbib\@empty
\bibitem [{\citenamefont {Guo}\ \emph {et~al.}(2023)\citenamefont {Guo}, \citenamefont {Miao}, \citenamefont {Huang}, \citenamefont {Lygo}, \citenamefont {Dai},\ and\ \citenamefont {Stemmer}}]{guo_2d_weyl_2023}%
  \BibitemOpen
  \bibfield  {author} {\bibinfo {author} {\bibfnamefont {B.}~\bibnamefont {Guo}}, \bibinfo {author} {\bibfnamefont {W.}~\bibnamefont {Miao}}, \bibinfo {author} {\bibfnamefont {V.}~\bibnamefont {Huang}}, \bibinfo {author} {\bibfnamefont {A.~C.}\ \bibnamefont {Lygo}}, \bibinfo {author} {\bibfnamefont {X.}~\bibnamefont {Dai}},\ and\ \bibinfo {author} {\bibfnamefont {S.}~\bibnamefont {Stemmer}},\ }\href {https://doi.org/10.1103/PhysRevLett.131.046601} {\bibfield  {journal} {\bibinfo  {journal} {Phys. Rev. Lett.}\ }\textbf {\bibinfo {volume} {131}},\ \bibinfo {pages} {046601} (\bibinfo {year} {2023})}\BibitemShut {NoStop}%
\bibitem [{\citenamefont {Guo}\ \emph {et~al.}(2022)\citenamefont {Guo}, \citenamefont {Lygo}, \citenamefont {Dai},\ and\ \citenamefont {Stemmer}}]{guo_2022}%
  \BibitemOpen
  \bibfield  {author} {\bibinfo {author} {\bibfnamefont {B.}~\bibnamefont {Guo}}, \bibinfo {author} {\bibfnamefont {A.~C.}\ \bibnamefont {Lygo}}, \bibinfo {author} {\bibfnamefont {X.}~\bibnamefont {Dai}},\ and\ \bibinfo {author} {\bibfnamefont {S.}~\bibnamefont {Stemmer}},\ }\href {https://doi.org/10.1063/5.0102703} {\bibfield  {journal} {\bibinfo  {journal} {APL Mater.}\ }\textbf {\bibinfo {volume} {10}},\ \bibinfo {pages} {091116} (\bibinfo {year} {2022})}\BibitemShut {NoStop}%
\bibitem [{\citenamefont {Lygo}\ \emph {et~al.}(2023)\citenamefont {Lygo}, \citenamefont {Guo}, \citenamefont {Rashidi}, \citenamefont {Huang}, \citenamefont {Cuadros-Romero},\ and\ \citenamefont {Stemmer}}]{lygo_two-dimensional_2023}%
  \BibitemOpen
  \bibfield  {author} {\bibinfo {author} {\bibfnamefont {A.~C.}\ \bibnamefont {Lygo}}, \bibinfo {author} {\bibfnamefont {B.}~\bibnamefont {Guo}}, \bibinfo {author} {\bibfnamefont {A.}~\bibnamefont {Rashidi}}, \bibinfo {author} {\bibfnamefont {V.}~\bibnamefont {Huang}}, \bibinfo {author} {\bibfnamefont {P.}~\bibnamefont {Cuadros-Romero}},\ and\ \bibinfo {author} {\bibfnamefont {S.}~\bibnamefont {Stemmer}},\ }\href {https://doi.org/10.1103/PhysRevLett.130.046201} {\bibfield  {journal} {\bibinfo  {journal} {Phys. Rev. Lett.}\ }\textbf {\bibinfo {volume} {130}},\ \bibinfo {pages} {046201} (\bibinfo {year} {2023})}\BibitemShut {NoStop}%
\bibitem [{\citenamefont {Wan}\ \emph {et~al.}(2011)\citenamefont {Wan}, \citenamefont {Turner}, \citenamefont {Vishwanath},\ and\ \citenamefont {Savrasov}}]{wan_weyl}%
  \BibitemOpen
  \bibfield  {author} {\bibinfo {author} {\bibfnamefont {X.}~\bibnamefont {Wan}}, \bibinfo {author} {\bibfnamefont {A.~M.}\ \bibnamefont {Turner}}, \bibinfo {author} {\bibfnamefont {A.}~\bibnamefont {Vishwanath}},\ and\ \bibinfo {author} {\bibfnamefont {S.~Y.}\ \bibnamefont {Savrasov}},\ }\href {https://doi.org/10.1103/PhysRevB.83.205101} {\bibfield  {journal} {\bibinfo  {journal} {Phys. Rev. B}\ }\textbf {\bibinfo {volume} {83}},\ \bibinfo {pages} {205101} (\bibinfo {year} {2011})}\BibitemShut {NoStop}%
\bibitem [{\citenamefont {Ruan}\ \emph {et~al.}(2016)\citenamefont {Ruan}, \citenamefont {Jian}, \citenamefont {Yao}, \citenamefont {Zhang}, \citenamefont {Zhang},\ and\ \citenamefont {Xing}}]{ruan2016symmetry}%
  \BibitemOpen
  \bibfield  {author} {\bibinfo {author} {\bibfnamefont {J.}~\bibnamefont {Ruan}}, \bibinfo {author} {\bibfnamefont {S.-K.}\ \bibnamefont {Jian}}, \bibinfo {author} {\bibfnamefont {H.}~\bibnamefont {Yao}}, \bibinfo {author} {\bibfnamefont {H.}~\bibnamefont {Zhang}}, \bibinfo {author} {\bibfnamefont {S.-C.}\ \bibnamefont {Zhang}},\ and\ \bibinfo {author} {\bibfnamefont {D.}~\bibnamefont {Xing}},\ }\href@noop {} {\bibfield  {journal} {\bibinfo  {journal} {Nature communications}\ }\textbf {\bibinfo {volume} {7}},\ \bibinfo {pages} {11136} (\bibinfo {year} {2016})}\BibitemShut {NoStop}%
\bibitem [{\citenamefont {Armitage}\ \emph {et~al.}(2018)\citenamefont {Armitage}, \citenamefont {Mele},\ and\ \citenamefont {Vishwanath}}]{armitage_weyl_2018}%
  \BibitemOpen
  \bibfield  {author} {\bibinfo {author} {\bibfnamefont {N.}~\bibnamefont {Armitage}}, \bibinfo {author} {\bibfnamefont {E.}~\bibnamefont {Mele}},\ and\ \bibinfo {author} {\bibfnamefont {A.}~\bibnamefont {Vishwanath}},\ }\href {https://doi.org/10.1103/RevModPhys.90.015001} {\bibfield  {journal} {\bibinfo  {journal} {Rev. Mod. Phys.}\ }\textbf {\bibinfo {volume} {90}},\ \bibinfo {pages} {015001} (\bibinfo {year} {2018})}\BibitemShut {NoStop}%
\bibitem [{\citenamefont {Lv}\ \emph {et~al.}(2015)\citenamefont {Lv}, \citenamefont {Weng}, \citenamefont {Fu}, \citenamefont {Wang}, \citenamefont {Miao}, \citenamefont {Ma}, \citenamefont {Richard}, \citenamefont {Huang}, \citenamefont {Zhao}, \citenamefont {Chen}, \citenamefont {Fang}, \citenamefont {Dai}, \citenamefont {Qian},\ and\ \citenamefont {Ding}}]{Lv_TaAs_exp}%
  \BibitemOpen
  \bibfield  {author} {\bibinfo {author} {\bibfnamefont {B.~Q.}\ \bibnamefont {Lv}}, \bibinfo {author} {\bibfnamefont {H.~M.}\ \bibnamefont {Weng}}, \bibinfo {author} {\bibfnamefont {B.~B.}\ \bibnamefont {Fu}}, \bibinfo {author} {\bibfnamefont {X.~P.}\ \bibnamefont {Wang}}, \bibinfo {author} {\bibfnamefont {H.}~\bibnamefont {Miao}}, \bibinfo {author} {\bibfnamefont {J.}~\bibnamefont {Ma}}, \bibinfo {author} {\bibfnamefont {P.}~\bibnamefont {Richard}}, \bibinfo {author} {\bibfnamefont {X.~C.}\ \bibnamefont {Huang}}, \bibinfo {author} {\bibfnamefont {L.~X.}\ \bibnamefont {Zhao}}, \bibinfo {author} {\bibfnamefont {G.~F.}\ \bibnamefont {Chen}}, \bibinfo {author} {\bibfnamefont {Z.}~\bibnamefont {Fang}}, \bibinfo {author} {\bibfnamefont {X.}~\bibnamefont {Dai}}, \bibinfo {author} {\bibfnamefont {T.}~\bibnamefont {Qian}},\ and\ \bibinfo {author} {\bibfnamefont {H.}~\bibnamefont {Ding}},\ }\href {https://doi.org/10.1103/PhysRevX.5.031013} {\bibfield  {journal} {\bibinfo  {journal} {Phys. Rev. X}\ }\textbf {\bibinfo
  {volume} {5}},\ \bibinfo {pages} {031013} (\bibinfo {year} {2015})}\BibitemShut {NoStop}%
\bibitem [{\citenamefont {Weng}\ \emph {et~al.}(2015)\citenamefont {Weng}, \citenamefont {Fang}, \citenamefont {Fang}, \citenamefont {Bernevig},\ and\ \citenamefont {Dai}}]{Weng_TaAs_theory}%
  \BibitemOpen
  \bibfield  {author} {\bibinfo {author} {\bibfnamefont {H.}~\bibnamefont {Weng}}, \bibinfo {author} {\bibfnamefont {C.}~\bibnamefont {Fang}}, \bibinfo {author} {\bibfnamefont {Z.}~\bibnamefont {Fang}}, \bibinfo {author} {\bibfnamefont {B.~A.}\ \bibnamefont {Bernevig}},\ and\ \bibinfo {author} {\bibfnamefont {X.}~\bibnamefont {Dai}},\ }\href {https://doi.org/10.1103/PhysRevX.5.011029} {\bibfield  {journal} {\bibinfo  {journal} {Phys. Rev. X}\ }\textbf {\bibinfo {volume} {5}},\ \bibinfo {pages} {011029} (\bibinfo {year} {2015})}\BibitemShut {NoStop}%
\bibitem [{\citenamefont {Xu}\ \emph {et~al.}(2015)\citenamefont {Xu}, \citenamefont {Belopolski}, \citenamefont {Alidoust}, \citenamefont {Neupane}, \citenamefont {Bian}, \citenamefont {Zhang}, \citenamefont {Sankar}, \citenamefont {Chang}, \citenamefont {Yuan}, \citenamefont {Lee} \emph {et~al.}}]{xu2015discovery}%
  \BibitemOpen
  \bibfield  {author} {\bibinfo {author} {\bibfnamefont {S.-Y.}\ \bibnamefont {Xu}}, \bibinfo {author} {\bibfnamefont {I.}~\bibnamefont {Belopolski}}, \bibinfo {author} {\bibfnamefont {N.}~\bibnamefont {Alidoust}}, \bibinfo {author} {\bibfnamefont {M.}~\bibnamefont {Neupane}}, \bibinfo {author} {\bibfnamefont {G.}~\bibnamefont {Bian}}, \bibinfo {author} {\bibfnamefont {C.}~\bibnamefont {Zhang}}, \bibinfo {author} {\bibfnamefont {R.}~\bibnamefont {Sankar}}, \bibinfo {author} {\bibfnamefont {G.}~\bibnamefont {Chang}}, \bibinfo {author} {\bibfnamefont {Z.}~\bibnamefont {Yuan}}, \bibinfo {author} {\bibfnamefont {C.-C.}\ \bibnamefont {Lee}}, \emph {et~al.},\ }\href@noop {} {\bibfield  {journal} {\bibinfo  {journal} {Science}\ }\textbf {\bibinfo {volume} {349}},\ \bibinfo {pages} {613} (\bibinfo {year} {2015})}\BibitemShut {NoStop}%
\bibitem [{\citenamefont {Xiang}\ \emph {et~al.}(2019)\citenamefont {Xiang}, \citenamefont {Hu}, \citenamefont {Song}, \citenamefont {Lv}, \citenamefont {Zhang}, \citenamefont {Zhao}, \citenamefont {Li}, \citenamefont {Chen}, \citenamefont {Zhang}, \citenamefont {Wang}, \citenamefont {Yang}, \citenamefont {Dai}, \citenamefont {Steglich}, \citenamefont {Chen},\ and\ \citenamefont {Sun}}]{chiral_zero_sound_exp}%
  \BibitemOpen
  \bibfield  {author} {\bibinfo {author} {\bibfnamefont {J.}~\bibnamefont {Xiang}}, \bibinfo {author} {\bibfnamefont {S.}~\bibnamefont {Hu}}, \bibinfo {author} {\bibfnamefont {Z.}~\bibnamefont {Song}}, \bibinfo {author} {\bibfnamefont {M.}~\bibnamefont {Lv}}, \bibinfo {author} {\bibfnamefont {J.}~\bibnamefont {Zhang}}, \bibinfo {author} {\bibfnamefont {L.}~\bibnamefont {Zhao}}, \bibinfo {author} {\bibfnamefont {W.}~\bibnamefont {Li}}, \bibinfo {author} {\bibfnamefont {Z.}~\bibnamefont {Chen}}, \bibinfo {author} {\bibfnamefont {S.}~\bibnamefont {Zhang}}, \bibinfo {author} {\bibfnamefont {J.-T.}\ \bibnamefont {Wang}}, \bibinfo {author} {\bibfnamefont {Y.-f.}\ \bibnamefont {Yang}}, \bibinfo {author} {\bibfnamefont {X.}~\bibnamefont {Dai}}, \bibinfo {author} {\bibfnamefont {F.}~\bibnamefont {Steglich}}, \bibinfo {author} {\bibfnamefont {G.}~\bibnamefont {Chen}},\ and\ \bibinfo {author} {\bibfnamefont {P.}~\bibnamefont {Sun}},\ }\href {https://doi.org/10.1103/PhysRevX.9.031036} {\bibfield  {journal} {\bibinfo
  {journal} {Phys. Rev. X}\ }\textbf {\bibinfo {volume} {9}},\ \bibinfo {pages} {031036} (\bibinfo {year} {2019})}\BibitemShut {NoStop}%
\bibitem [{\citenamefont {Song}\ and\ \citenamefont {Dai}(2019)}]{chiral_zero_sound_song}%
  \BibitemOpen
  \bibfield  {author} {\bibinfo {author} {\bibfnamefont {Z.}~\bibnamefont {Song}}\ and\ \bibinfo {author} {\bibfnamefont {X.}~\bibnamefont {Dai}},\ }\href {https://doi.org/10.1103/PhysRevX.9.021053} {\bibfield  {journal} {\bibinfo  {journal} {Phys. Rev. X}\ }\textbf {\bibinfo {volume} {9}},\ \bibinfo {pages} {021053} (\bibinfo {year} {2019})}\BibitemShut {NoStop}%
\bibitem [{\citenamefont {Cano}\ \emph {et~al.}(2017)\citenamefont {Cano}, \citenamefont {Bradlyn}, \citenamefont {Wang}, \citenamefont {Hirschberger}, \citenamefont {Ong},\ and\ \citenamefont {Bernevig}}]{cano_chiral_2017}%
  \BibitemOpen
  \bibfield  {author} {\bibinfo {author} {\bibfnamefont {J.}~\bibnamefont {Cano}}, \bibinfo {author} {\bibfnamefont {B.}~\bibnamefont {Bradlyn}}, \bibinfo {author} {\bibfnamefont {Z.}~\bibnamefont {Wang}}, \bibinfo {author} {\bibfnamefont {M.}~\bibnamefont {Hirschberger}}, \bibinfo {author} {\bibfnamefont {N.~P.}\ \bibnamefont {Ong}},\ and\ \bibinfo {author} {\bibfnamefont {B.~A.}\ \bibnamefont {Bernevig}},\ }\href {https://doi.org/10.1103/PhysRevB.95.161306} {\bibfield  {journal} {\bibinfo  {journal} {Phys. Rev. B}\ }\textbf {\bibinfo {volume} {95}},\ \bibinfo {pages} {161306} (\bibinfo {year} {2017})}\BibitemShut {NoStop}%
\bibitem [{\citenamefont {Baidya}\ and\ \citenamefont {Vanderbilt}(2020)}]{baidya_cd3as2_zeeman_prb}%
  \BibitemOpen
  \bibfield  {author} {\bibinfo {author} {\bibfnamefont {S.}~\bibnamefont {Baidya}}\ and\ \bibinfo {author} {\bibfnamefont {D.}~\bibnamefont {Vanderbilt}},\ }\href {https://doi.org/10.1103/PhysRevB.102.165115} {\bibfield  {journal} {\bibinfo  {journal} {Phys. Rev. B}\ }\textbf {\bibinfo {volume} {102}},\ \bibinfo {pages} {165115} (\bibinfo {year} {2020})}\BibitemShut {NoStop}%
\bibitem [{\citenamefont {Xiao}\ \emph {et~al.}(2022)\citenamefont {Xiao}, \citenamefont {Held}, \citenamefont {Rable}, \citenamefont {Ghosh}, \citenamefont {Wang}, \citenamefont {Mkhoyan},\ and\ \citenamefont {Samarth}}]{xiao_cd3as2_mag_doping}%
  \BibitemOpen
  \bibfield  {author} {\bibinfo {author} {\bibfnamefont {R.}~\bibnamefont {Xiao}}, \bibinfo {author} {\bibfnamefont {J.~T.}\ \bibnamefont {Held}}, \bibinfo {author} {\bibfnamefont {J.}~\bibnamefont {Rable}}, \bibinfo {author} {\bibfnamefont {S.}~\bibnamefont {Ghosh}}, \bibinfo {author} {\bibfnamefont {K.}~\bibnamefont {Wang}}, \bibinfo {author} {\bibfnamefont {K.~A.}\ \bibnamefont {Mkhoyan}},\ and\ \bibinfo {author} {\bibfnamefont {N.}~\bibnamefont {Samarth}},\ }\href {https://doi.org/10.1103/PhysRevMaterials.6.024203} {\bibfield  {journal} {\bibinfo  {journal} {Phys. Rev. Mater.}\ }\textbf {\bibinfo {volume} {6}},\ \bibinfo {pages} {024203} (\bibinfo {year} {2022})}\BibitemShut {NoStop}%
\bibitem [{\citenamefont {You}\ \emph {et~al.}(2019)\citenamefont {You}, \citenamefont {Chen}, \citenamefont {Zhang}, \citenamefont {Sheng}, \citenamefont {Yang},\ and\ \citenamefont {Su}}]{you_two-dimensional_2019}%
  \BibitemOpen
  \bibfield  {author} {\bibinfo {author} {\bibfnamefont {J.-Y.}\ \bibnamefont {You}}, \bibinfo {author} {\bibfnamefont {C.}~\bibnamefont {Chen}}, \bibinfo {author} {\bibfnamefont {Z.}~\bibnamefont {Zhang}}, \bibinfo {author} {\bibfnamefont {X.-L.}\ \bibnamefont {Sheng}}, \bibinfo {author} {\bibfnamefont {S.~A.}\ \bibnamefont {Yang}},\ and\ \bibinfo {author} {\bibfnamefont {G.}~\bibnamefont {Su}},\ }\href {https://doi.org/10.1103/PhysRevB.100.064408} {\bibfield  {journal} {\bibinfo  {journal} {Phys. Rev. B}\ }\textbf {\bibinfo {volume} {100}},\ \bibinfo {pages} {064408} (\bibinfo {year} {2019})}\BibitemShut {NoStop}%
\bibitem [{\citenamefont {Li}\ \emph {et~al.}(2019)\citenamefont {Li}, \citenamefont {Li}, \citenamefont {Du}, \citenamefont {Wang}, \citenamefont {Gu}, \citenamefont {Zhang}, \citenamefont {He}, \citenamefont {Duan},\ and\ \citenamefont {Xu}}]{MBT_weyl}%
  \BibitemOpen
  \bibfield  {author} {\bibinfo {author} {\bibfnamefont {J.}~\bibnamefont {Li}}, \bibinfo {author} {\bibfnamefont {Y.}~\bibnamefont {Li}}, \bibinfo {author} {\bibfnamefont {S.}~\bibnamefont {Du}}, \bibinfo {author} {\bibfnamefont {Z.}~\bibnamefont {Wang}}, \bibinfo {author} {\bibfnamefont {B.-L.}\ \bibnamefont {Gu}}, \bibinfo {author} {\bibfnamefont {S.-C.}\ \bibnamefont {Zhang}}, \bibinfo {author} {\bibfnamefont {K.}~\bibnamefont {He}}, \bibinfo {author} {\bibfnamefont {W.}~\bibnamefont {Duan}},\ and\ \bibinfo {author} {\bibfnamefont {Y.}~\bibnamefont {Xu}},\ }\href {https://doi.org/10.1126/sciadv.aaw5685} {\bibfield  {journal} {\bibinfo  {journal} {Science Advances}\ }\textbf {\bibinfo {volume} {5}},\ \bibinfo {pages} {eaaw5685} (\bibinfo {year} {2019})}\BibitemShut {NoStop}%
\bibitem [{\citenamefont {Wang}\ \emph {et~al.}(2019)\citenamefont {Wang}, \citenamefont {Jo}, \citenamefont {Kuthanazhi}, \citenamefont {Wu}, \citenamefont {McQueeney}, \citenamefont {Kaminski},\ and\ \citenamefont {Canfield}}]{single_weyl_Eucd2as2}%
  \BibitemOpen
  \bibfield  {author} {\bibinfo {author} {\bibfnamefont {L.-L.}\ \bibnamefont {Wang}}, \bibinfo {author} {\bibfnamefont {N.~H.}\ \bibnamefont {Jo}}, \bibinfo {author} {\bibfnamefont {B.}~\bibnamefont {Kuthanazhi}}, \bibinfo {author} {\bibfnamefont {Y.}~\bibnamefont {Wu}}, \bibinfo {author} {\bibfnamefont {R.~J.}\ \bibnamefont {McQueeney}}, \bibinfo {author} {\bibfnamefont {A.}~\bibnamefont {Kaminski}},\ and\ \bibinfo {author} {\bibfnamefont {P.~C.}\ \bibnamefont {Canfield}},\ }\href {https://doi.org/10.1103/PhysRevB.99.245147} {\bibfield  {journal} {\bibinfo  {journal} {Phys. Rev. B}\ }\textbf {\bibinfo {volume} {99}},\ \bibinfo {pages} {245147} (\bibinfo {year} {2019})}\BibitemShut {NoStop}%
\bibitem [{\citenamefont {Zhao}\ \emph {et~al.}(2022)\citenamefont {Zhao}, \citenamefont {Ma}, \citenamefont {Guo},\ and\ \citenamefont {Lu}}]{zhao2022two}%
  \BibitemOpen
  \bibfield  {author} {\bibinfo {author} {\bibfnamefont {X.}~\bibnamefont {Zhao}}, \bibinfo {author} {\bibfnamefont {F.}~\bibnamefont {Ma}}, \bibinfo {author} {\bibfnamefont {P.-J.}\ \bibnamefont {Guo}},\ and\ \bibinfo {author} {\bibfnamefont {Z.-Y.}\ \bibnamefont {Lu}},\ }\href@noop {} {\bibfield  {journal} {\bibinfo  {journal} {Physical Review Research}\ }\textbf {\bibinfo {volume} {4}},\ \bibinfo {pages} {043183} (\bibinfo {year} {2022})}\BibitemShut {NoStop}%
\bibitem [{\citenamefont {Nie}\ \emph {et~al.}(2022)\citenamefont {Nie}, \citenamefont {Hashimoto},\ and\ \citenamefont {Prinz}}]{single_mag_weyl}%
  \BibitemOpen
  \bibfield  {author} {\bibinfo {author} {\bibfnamefont {S.}~\bibnamefont {Nie}}, \bibinfo {author} {\bibfnamefont {T.}~\bibnamefont {Hashimoto}},\ and\ \bibinfo {author} {\bibfnamefont {F.~B.}\ \bibnamefont {Prinz}},\ }\href {https://doi.org/10.1103/PhysRevLett.128.176401} {\bibfield  {journal} {\bibinfo  {journal} {Phys. Rev. Lett.}\ }\textbf {\bibinfo {volume} {128}},\ \bibinfo {pages} {176401} (\bibinfo {year} {2022})}\BibitemShut {NoStop}%
\bibitem [{\citenamefont {Wang}\ \emph {et~al.}(2022)\citenamefont {Wang}, \citenamefont {Zhou}, \citenamefont {Zhang}, \citenamefont {Wu}, \citenamefont {Yu},\ and\ \citenamefont {Yang}}]{weyl_nonmag}%
  \BibitemOpen
  \bibfield  {author} {\bibinfo {author} {\bibfnamefont {X.}~\bibnamefont {Wang}}, \bibinfo {author} {\bibfnamefont {F.}~\bibnamefont {Zhou}}, \bibinfo {author} {\bibfnamefont {Z.}~\bibnamefont {Zhang}}, \bibinfo {author} {\bibfnamefont {W.}~\bibnamefont {Wu}}, \bibinfo {author} {\bibfnamefont {Z.-M.}\ \bibnamefont {Yu}},\ and\ \bibinfo {author} {\bibfnamefont {S.~A.}\ \bibnamefont {Yang}},\ }\href {https://doi.org/10.1103/PhysRevB.106.195129} {\bibfield  {journal} {\bibinfo  {journal} {Phys. Rev. B}\ }\textbf {\bibinfo {volume} {106}},\ \bibinfo {pages} {195129} (\bibinfo {year} {2022})}\BibitemShut {NoStop}%
\bibitem [{\citenamefont {Xiang}\ and\ \citenamefont {Wang}(2024)}]{PhysRevB.109.075419}%
  \BibitemOpen
  \bibfield  {author} {\bibinfo {author} {\bibfnamefont {L.}~\bibnamefont {Xiang}}\ and\ \bibinfo {author} {\bibfnamefont {J.}~\bibnamefont {Wang}},\ }\href {https://doi.org/10.1103/PhysRevB.109.075419} {\bibfield  {journal} {\bibinfo  {journal} {Phys. Rev. B}\ }\textbf {\bibinfo {volume} {109}},\ \bibinfo {pages} {075419} (\bibinfo {year} {2024})}\BibitemShut {NoStop}%
\bibitem [{\citenamefont {Fang}\ and\ \citenamefont {Fu}(2015)}]{fang_new_2015}%
  \BibitemOpen
  \bibfield  {author} {\bibinfo {author} {\bibfnamefont {C.}~\bibnamefont {Fang}}\ and\ \bibinfo {author} {\bibfnamefont {L.}~\bibnamefont {Fu}},\ }\href {https://doi.org/10.1103/PhysRevB.91.161105} {\bibfield  {journal} {\bibinfo  {journal} {Phys. Rev. B}\ }\textbf {\bibinfo {volume} {91}},\ \bibinfo {pages} {161105} (\bibinfo {year} {2015})}\BibitemShut {NoStop}%
\bibitem [{\citenamefont {Ahn}\ and\ \citenamefont {Yang}(2017)}]{ahn_unconventional_2017}%
  \BibitemOpen
  \bibfield  {author} {\bibinfo {author} {\bibfnamefont {J.}~\bibnamefont {Ahn}}\ and\ \bibinfo {author} {\bibfnamefont {B.-J.}\ \bibnamefont {Yang}},\ }\href {https://doi.org/10.1103/PhysRevLett.118.156401} {\bibfield  {journal} {\bibinfo  {journal} {Phys. Rev. Lett.}\ }\textbf {\bibinfo {volume} {118}},\ \bibinfo {pages} {156401} (\bibinfo {year} {2017})}\BibitemShut {NoStop}%
\bibitem [{\citenamefont {Zyuzin}\ \emph {et~al.}(2011)\citenamefont {Zyuzin}, \citenamefont {Hook},\ and\ \citenamefont {Burkov}}]{burkov_phase_tran}%
  \BibitemOpen
  \bibfield  {author} {\bibinfo {author} {\bibfnamefont {A.~A.}\ \bibnamefont {Zyuzin}}, \bibinfo {author} {\bibfnamefont {M.~D.}\ \bibnamefont {Hook}},\ and\ \bibinfo {author} {\bibfnamefont {A.~A.}\ \bibnamefont {Burkov}},\ }\href {https://doi.org/10.1103/PhysRevB.83.245428} {\bibfield  {journal} {\bibinfo  {journal} {Phys. Rev. B}\ }\textbf {\bibinfo {volume} {83}},\ \bibinfo {pages} {245428} (\bibinfo {year} {2011})}\BibitemShut {NoStop}%
\bibitem [{\citenamefont {Jiang}\ \emph {et~al.}(2023)\citenamefont {Jiang}, \citenamefont {Ermolaev}, \citenamefont {Moon}, \citenamefont {Kipshidze}, \citenamefont {Belenky}, \citenamefont {Svensson}, \citenamefont {Ozerov}, \citenamefont {Smirnov}, \citenamefont {Jiang},\ and\ \citenamefont {Suchalkin}}]{gfactor_eng_prb_23}%
  \BibitemOpen
  \bibfield  {author} {\bibinfo {author} {\bibfnamefont {Y.}~\bibnamefont {Jiang}}, \bibinfo {author} {\bibfnamefont {M.}~\bibnamefont {Ermolaev}}, \bibinfo {author} {\bibfnamefont {S.}~\bibnamefont {Moon}}, \bibinfo {author} {\bibfnamefont {G.}~\bibnamefont {Kipshidze}}, \bibinfo {author} {\bibfnamefont {G.}~\bibnamefont {Belenky}}, \bibinfo {author} {\bibfnamefont {S.}~\bibnamefont {Svensson}}, \bibinfo {author} {\bibfnamefont {M.}~\bibnamefont {Ozerov}}, \bibinfo {author} {\bibfnamefont {D.}~\bibnamefont {Smirnov}}, \bibinfo {author} {\bibfnamefont {Z.}~\bibnamefont {Jiang}},\ and\ \bibinfo {author} {\bibfnamefont {S.}~\bibnamefont {Suchalkin}},\ }\href {https://doi.org/10.1103/PhysRevB.108.L121201} {\bibfield  {journal} {\bibinfo  {journal} {Phys. Rev. B}\ }\textbf {\bibinfo {volume} {108}},\ \bibinfo {pages} {L121201} (\bibinfo {year} {2023})}\BibitemShut {NoStop}%
\bibitem [{\citenamefont {Liu}\ \emph {et~al.}(2013)\citenamefont {Liu}, \citenamefont {Hsu},\ and\ \citenamefont {Liu}}]{liu_-plane_2013}%
  \BibitemOpen
  \bibfield  {author} {\bibinfo {author} {\bibfnamefont {X.}~\bibnamefont {Liu}}, \bibinfo {author} {\bibfnamefont {H.-C.}\ \bibnamefont {Hsu}},\ and\ \bibinfo {author} {\bibfnamefont {C.-X.}\ \bibnamefont {Liu}},\ }\href {https://doi.org/10.1103/PhysRevLett.111.086802} {\bibfield  {journal} {\bibinfo  {journal} {Phys. Rev. Lett.}\ }\textbf {\bibinfo {volume} {111}},\ \bibinfo {pages} {086802} (\bibinfo {year} {2013})}\BibitemShut {NoStop}%
\bibitem [{\citenamefont {Sun}\ \emph {et~al.}(2022)\citenamefont {Sun}, \citenamefont {Weng},\ and\ \citenamefont {Dai}}]{sun_possible_2022}%
  \BibitemOpen
  \bibfield  {author} {\bibinfo {author} {\bibfnamefont {S.}~\bibnamefont {Sun}}, \bibinfo {author} {\bibfnamefont {H.}~\bibnamefont {Weng}},\ and\ \bibinfo {author} {\bibfnamefont {X.}~\bibnamefont {Dai}},\ }\href {https://doi.org/10.1103/PhysRevB.106.L241105} {\bibfield  {journal} {\bibinfo  {journal} {Phys. Rev. B}\ }\textbf {\bibinfo {volume} {106}},\ \bibinfo {pages} {L241105} (\bibinfo {year} {2022})}\BibitemShut {NoStop}%
\bibitem [{\citenamefont {Cao}\ \emph {et~al.}(2023)\citenamefont {Cao}, \citenamefont {Jiang}, \citenamefont {Li}, \citenamefont {Tu}, \citenamefont {Zhou}, \citenamefont {Zhou},\ and\ \citenamefont {Yao}}]{yao_inplane_qah}%
  \BibitemOpen
  \bibfield  {author} {\bibinfo {author} {\bibfnamefont {J.}~\bibnamefont {Cao}}, \bibinfo {author} {\bibfnamefont {W.}~\bibnamefont {Jiang}}, \bibinfo {author} {\bibfnamefont {X.-P.}\ \bibnamefont {Li}}, \bibinfo {author} {\bibfnamefont {D.}~\bibnamefont {Tu}}, \bibinfo {author} {\bibfnamefont {J.}~\bibnamefont {Zhou}}, \bibinfo {author} {\bibfnamefont {J.}~\bibnamefont {Zhou}},\ and\ \bibinfo {author} {\bibfnamefont {Y.}~\bibnamefont {Yao}},\ }\href {https://doi.org/10.1103/PhysRevLett.130.166702} {\bibfield  {journal} {\bibinfo  {journal} {Phys. Rev. Lett.}\ }\textbf {\bibinfo {volume} {130}},\ \bibinfo {pages} {166702} (\bibinfo {year} {2023})}\BibitemShut {NoStop}%
\bibitem [{\citenamefont {Kurumaji}(2023)}]{kurumaji_iphall_symm}%
  \BibitemOpen
  \bibfield  {author} {\bibinfo {author} {\bibfnamefont {T.}~\bibnamefont {Kurumaji}},\ }\href {https://doi.org/10.1103/PhysRevResearch.5.023138} {\bibfield  {journal} {\bibinfo  {journal} {Phys. Rev. Res.}\ }\textbf {\bibinfo {volume} {5}},\ \bibinfo {pages} {023138} (\bibinfo {year} {2023})}\BibitemShut {NoStop}%
\bibitem [{\citenamefont {Cullen}\ \emph {et~al.}(2021)\citenamefont {Cullen}, \citenamefont {Bhalla}, \citenamefont {Marcellina}, \citenamefont {Hamilton},\ and\ \citenamefont {Culcer}}]{ipahe_prl_21}%
  \BibitemOpen
  \bibfield  {author} {\bibinfo {author} {\bibfnamefont {J.~H.}\ \bibnamefont {Cullen}}, \bibinfo {author} {\bibfnamefont {P.}~\bibnamefont {Bhalla}}, \bibinfo {author} {\bibfnamefont {E.}~\bibnamefont {Marcellina}}, \bibinfo {author} {\bibfnamefont {A.~R.}\ \bibnamefont {Hamilton}},\ and\ \bibinfo {author} {\bibfnamefont {D.}~\bibnamefont {Culcer}},\ }\href {https://doi.org/10.1103/PhysRevLett.126.256601} {\bibfield  {journal} {\bibinfo  {journal} {Phys. Rev. Lett.}\ }\textbf {\bibinfo {volume} {126}},\ \bibinfo {pages} {256601} (\bibinfo {year} {2021})}\BibitemShut {NoStop}%
\bibitem [{\citenamefont {Ren}\ \emph {et~al.}(2016)\citenamefont {Ren}, \citenamefont {Zeng}, \citenamefont {Deng}, \citenamefont {Yang}, \citenamefont {Pan},\ and\ \citenamefont {Qiao}}]{ren_ipqah_prb_17}%
  \BibitemOpen
  \bibfield  {author} {\bibinfo {author} {\bibfnamefont {Y.}~\bibnamefont {Ren}}, \bibinfo {author} {\bibfnamefont {J.}~\bibnamefont {Zeng}}, \bibinfo {author} {\bibfnamefont {X.}~\bibnamefont {Deng}}, \bibinfo {author} {\bibfnamefont {F.}~\bibnamefont {Yang}}, \bibinfo {author} {\bibfnamefont {H.}~\bibnamefont {Pan}},\ and\ \bibinfo {author} {\bibfnamefont {Z.}~\bibnamefont {Qiao}},\ }\href {https://doi.org/10.1103/PhysRevB.94.085411} {\bibfield  {journal} {\bibinfo  {journal} {Phys. Rev. B}\ }\textbf {\bibinfo {volume} {94}},\ \bibinfo {pages} {085411} (\bibinfo {year} {2016})}\BibitemShut {NoStop}%
\bibitem [{\citenamefont {Zhang}\ \emph {et~al.}(2019)\citenamefont {Zhang}, \citenamefont {Liu},\ and\ \citenamefont {Wang}}]{ipahe_jwang_prb_19}%
  \BibitemOpen
  \bibfield  {author} {\bibinfo {author} {\bibfnamefont {J.}~\bibnamefont {Zhang}}, \bibinfo {author} {\bibfnamefont {Z.}~\bibnamefont {Liu}},\ and\ \bibinfo {author} {\bibfnamefont {J.}~\bibnamefont {Wang}},\ }\href {https://doi.org/10.1103/PhysRevB.100.165117} {\bibfield  {journal} {\bibinfo  {journal} {Phys. Rev. B}\ }\textbf {\bibinfo {volume} {100}},\ \bibinfo {pages} {165117} (\bibinfo {year} {2019})}\BibitemShut {NoStop}%
\bibitem [{\citenamefont {Li}\ \emph {et~al.}(2023{\natexlab{a}})\citenamefont {Li}, \citenamefont {Wang}, \citenamefont {Li},\ and\ \citenamefont {Zhou}}]{ding_ipahe_arxiv}%
  \BibitemOpen
  \bibfield  {author} {\bibinfo {author} {\bibfnamefont {D.}~\bibnamefont {Li}}, \bibinfo {author} {\bibfnamefont {M.}~\bibnamefont {Wang}}, \bibinfo {author} {\bibfnamefont {D.}~\bibnamefont {Li}},\ and\ \bibinfo {author} {\bibfnamefont {J.}~\bibnamefont {Zhou}},\ }\href@noop {} {\bibinfo {title} {Switchable in-plane anomalous hall effect by magnetization orientation in monolayer $\mathrm{Mn}_{3}\mathrm{Si}_{2}\mathrm{Te}_{6}$}} (\bibinfo {year} {2023}{\natexlab{a}}),\ \Eprint {https://arxiv.org/abs/arXiv:2308.15343} {arXiv:2308.15343} \BibitemShut {NoStop}%
\bibitem [{Note1()}]{Note1}%
  \BibitemOpen
  \bibinfo {note} {We find that the term `in-plane anomalous Hall effect' is more appropriate for the phenomena described in this paper, as discussed in Ref.~\cite {kurumaji_iphall_symm}. However, it's worth noting that some papers refer to this as the `anomalous planar Hall effect' \cite {anomalous_planar_hall_prr, absence_aphe}.}\BibitemShut {Stop}%
\bibitem [{\citenamefont {Wang}\ \emph {et~al.}(2017)\citenamefont {Wang}, \citenamefont {Sun}, \citenamefont {Lu},\ and\ \citenamefont {Xie}}]{3d_quantum_hall_lu}%
  \BibitemOpen
  \bibfield  {author} {\bibinfo {author} {\bibfnamefont {C.~M.}\ \bibnamefont {Wang}}, \bibinfo {author} {\bibfnamefont {H.-P.}\ \bibnamefont {Sun}}, \bibinfo {author} {\bibfnamefont {H.-Z.}\ \bibnamefont {Lu}},\ and\ \bibinfo {author} {\bibfnamefont {X.~C.}\ \bibnamefont {Xie}},\ }\href {https://doi.org/10.1103/PhysRevLett.119.136806} {\bibfield  {journal} {\bibinfo  {journal} {Phys. Rev. Lett.}\ }\textbf {\bibinfo {volume} {119}},\ \bibinfo {pages} {136806} (\bibinfo {year} {2017})}\BibitemShut {NoStop}%
\bibitem [{\citenamefont {Chen}\ \emph {et~al.}(2021)\citenamefont {Chen}, \citenamefont {Liu}, \citenamefont {Wang}, \citenamefont {Lu},\ and\ \citenamefont {Xie}}]{hinge_dirac_lu}%
  \BibitemOpen
  \bibfield  {author} {\bibinfo {author} {\bibfnamefont {R.}~\bibnamefont {Chen}}, \bibinfo {author} {\bibfnamefont {T.}~\bibnamefont {Liu}}, \bibinfo {author} {\bibfnamefont {C.~M.}\ \bibnamefont {Wang}}, \bibinfo {author} {\bibfnamefont {H.-Z.}\ \bibnamefont {Lu}},\ and\ \bibinfo {author} {\bibfnamefont {X.~C.}\ \bibnamefont {Xie}},\ }\href {https://doi.org/10.1103/PhysRevLett.127.066801} {\bibfield  {journal} {\bibinfo  {journal} {Phys. Rev. Lett.}\ }\textbf {\bibinfo {volume} {127}},\ \bibinfo {pages} {066801} (\bibinfo {year} {2021})}\BibitemShut {NoStop}%
\bibitem [{\citenamefont {Xiao}\ \emph {et~al.}(2010)\citenamefont {Xiao}, \citenamefont {Chang},\ and\ \citenamefont {Niu}}]{rmp_niu}%
  \BibitemOpen
  \bibfield  {author} {\bibinfo {author} {\bibfnamefont {D.}~\bibnamefont {Xiao}}, \bibinfo {author} {\bibfnamefont {M.-C.}\ \bibnamefont {Chang}},\ and\ \bibinfo {author} {\bibfnamefont {Q.}~\bibnamefont {Niu}},\ }\href {https://doi.org/10.1103/RevModPhys.82.1959} {\bibfield  {journal} {\bibinfo  {journal} {Rev. Mod. Phys.}\ }\textbf {\bibinfo {volume} {82}},\ \bibinfo {pages} {1959} (\bibinfo {year} {2010})}\BibitemShut {NoStop}%
\bibitem [{\citenamefont {Nandy}\ \emph {et~al.}(2017)\citenamefont {Nandy}, \citenamefont {Sharma}, \citenamefont {Taraphder},\ and\ \citenamefont {Tewari}}]{chiral_anomaly_planar_hall_prl}%
  \BibitemOpen
  \bibfield  {author} {\bibinfo {author} {\bibfnamefont {S.}~\bibnamefont {Nandy}}, \bibinfo {author} {\bibfnamefont {G.}~\bibnamefont {Sharma}}, \bibinfo {author} {\bibfnamefont {A.}~\bibnamefont {Taraphder}},\ and\ \bibinfo {author} {\bibfnamefont {S.}~\bibnamefont {Tewari}},\ }\href {https://doi.org/10.1103/PhysRevLett.119.176804} {\bibfield  {journal} {\bibinfo  {journal} {Phys. Rev. Lett.}\ }\textbf {\bibinfo {volume} {119}},\ \bibinfo {pages} {176804} (\bibinfo {year} {2017})}\BibitemShut {NoStop}%
\bibitem [{\citenamefont {Wei}\ \emph {et~al.}(2023)\citenamefont {Wei}, \citenamefont {Feng},\ and\ \citenamefont {Weng}}]{weng_planr_hall}%
  \BibitemOpen
  \bibfield  {author} {\bibinfo {author} {\bibfnamefont {Y.-W.}\ \bibnamefont {Wei}}, \bibinfo {author} {\bibfnamefont {J.}~\bibnamefont {Feng}},\ and\ \bibinfo {author} {\bibfnamefont {H.}~\bibnamefont {Weng}},\ }\href {https://doi.org/10.1103/PhysRevB.107.075131} {\bibfield  {journal} {\bibinfo  {journal} {Phys. Rev. B}\ }\textbf {\bibinfo {volume} {107}},\ \bibinfo {pages} {075131} (\bibinfo {year} {2023})}\BibitemShut {NoStop}%
\bibitem [{\citenamefont {Li}\ \emph {et~al.}(2023{\natexlab{b}})\citenamefont {Li}, \citenamefont {Cao}, \citenamefont {Cui}, \citenamefont {Yu},\ and\ \citenamefont {Yao}}]{yao_planar_hall_prb_23}%
  \BibitemOpen
  \bibfield  {author} {\bibinfo {author} {\bibfnamefont {L.}~\bibnamefont {Li}}, \bibinfo {author} {\bibfnamefont {J.}~\bibnamefont {Cao}}, \bibinfo {author} {\bibfnamefont {C.}~\bibnamefont {Cui}}, \bibinfo {author} {\bibfnamefont {Z.-M.}\ \bibnamefont {Yu}},\ and\ \bibinfo {author} {\bibfnamefont {Y.}~\bibnamefont {Yao}},\ }\href {https://doi.org/10.1103/PhysRevB.108.085120} {\bibfield  {journal} {\bibinfo  {journal} {Phys. Rev. B}\ }\textbf {\bibinfo {volume} {108}},\ \bibinfo {pages} {085120} (\bibinfo {year} {2023}{\natexlab{b}})}\BibitemShut {NoStop}%
\bibitem [{\citenamefont {Song}\ \emph {et~al.}(2020)\citenamefont {Song}, \citenamefont {Sun}, \citenamefont {Xu}, \citenamefont {Nie}, \citenamefont {Weng}, \citenamefont {Fang},\ and\ \citenamefont {Dai}}]{song_chapter_2020}%
  \BibitemOpen
  \bibfield  {author} {\bibinfo {author} {\bibfnamefont {Z.}~\bibnamefont {Song}}, \bibinfo {author} {\bibfnamefont {S.}~\bibnamefont {Sun}}, \bibinfo {author} {\bibfnamefont {Y.}~\bibnamefont {Xu}}, \bibinfo {author} {\bibfnamefont {S.}~\bibnamefont {Nie}}, \bibinfo {author} {\bibfnamefont {H.}~\bibnamefont {Weng}}, \bibinfo {author} {\bibfnamefont {Z.}~\bibnamefont {Fang}},\ and\ \bibinfo {author} {\bibfnamefont {X.}~\bibnamefont {Dai}},\ }in\ \href {https://doi.org/10.1142/9789811231711_0013} {\emph {\bibinfo {booktitle} {Memorial {Volume} for {Shoucheng} {Zhang}}}}\ (\bibinfo  {publisher} {World Scientific},\ \bibinfo {year} {2020})\ pp.\ \bibinfo {pages} {263--281}\BibitemShut {NoStop}%
\bibitem [{\citenamefont {Sun}\ \emph {et~al.}(2020)\citenamefont {Sun}, \citenamefont {Song}, \citenamefont {Weng},\ and\ \citenamefont {Dai}}]{sun_topological_2020}%
  \BibitemOpen
  \bibfield  {author} {\bibinfo {author} {\bibfnamefont {S.}~\bibnamefont {Sun}}, \bibinfo {author} {\bibfnamefont {Z.}~\bibnamefont {Song}}, \bibinfo {author} {\bibfnamefont {H.}~\bibnamefont {Weng}},\ and\ \bibinfo {author} {\bibfnamefont {X.}~\bibnamefont {Dai}},\ }\href {https://doi.org/10.1103/PhysRevB.101.125118} {\bibfield  {journal} {\bibinfo  {journal} {Phys. Rev. B}\ }\textbf {\bibinfo {volume} {101}},\ \bibinfo {pages} {125118} (\bibinfo {year} {2020})}\BibitemShut {NoStop}%
\bibitem [{\citenamefont {Gao}\ \emph {et~al.}(2014)\citenamefont {Gao}, \citenamefont {Yang},\ and\ \citenamefont {Niu}}]{yang_gao_prl_bcp}%
  \BibitemOpen
  \bibfield  {author} {\bibinfo {author} {\bibfnamefont {Y.}~\bibnamefont {Gao}}, \bibinfo {author} {\bibfnamefont {S.~A.}\ \bibnamefont {Yang}},\ and\ \bibinfo {author} {\bibfnamefont {Q.}~\bibnamefont {Niu}},\ }\href {https://doi.org/10.1103/PhysRevLett.112.166601} {\bibfield  {journal} {\bibinfo  {journal} {Phys. Rev. Lett.}\ }\textbf {\bibinfo {volume} {112}},\ \bibinfo {pages} {166601} (\bibinfo {year} {2014})}\BibitemShut {NoStop}%
\bibitem [{\citenamefont {Huang}\ \emph {et~al.}(2023{\natexlab{a}})\citenamefont {Huang}, \citenamefont {Wang}, \citenamefont {Wang}, \citenamefont {Xiao}, \citenamefont {Li},\ and\ \citenamefont {Yang}}]{non_linear_inplane_mag_prb}%
  \BibitemOpen
  \bibfield  {author} {\bibinfo {author} {\bibfnamefont {Y.-X.}\ \bibnamefont {Huang}}, \bibinfo {author} {\bibfnamefont {Y.}~\bibnamefont {Wang}}, \bibinfo {author} {\bibfnamefont {H.}~\bibnamefont {Wang}}, \bibinfo {author} {\bibfnamefont {C.}~\bibnamefont {Xiao}}, \bibinfo {author} {\bibfnamefont {X.}~\bibnamefont {Li}},\ and\ \bibinfo {author} {\bibfnamefont {S.~A.}\ \bibnamefont {Yang}},\ }\href {https://doi.org/10.1103/PhysRevB.108.075155} {\bibfield  {journal} {\bibinfo  {journal} {Phys. Rev. B}\ }\textbf {\bibinfo {volume} {108}},\ \bibinfo {pages} {075155} (\bibinfo {year} {2023}{\natexlab{a}})}\BibitemShut {NoStop}%
\bibitem [{\citenamefont {Wang}\ \emph {et~al.}(2024)\citenamefont {Wang}, \citenamefont {Huang}, \citenamefont {Liu}, \citenamefont {Feng}, \citenamefont {Zhu}, \citenamefont {Wu}, \citenamefont {Xiao},\ and\ \citenamefont {Yang}}]{IPHE_orbital_prl}%
  \BibitemOpen
  \bibfield  {author} {\bibinfo {author} {\bibfnamefont {H.}~\bibnamefont {Wang}}, \bibinfo {author} {\bibfnamefont {Y.-X.}\ \bibnamefont {Huang}}, \bibinfo {author} {\bibfnamefont {H.}~\bibnamefont {Liu}}, \bibinfo {author} {\bibfnamefont {X.}~\bibnamefont {Feng}}, \bibinfo {author} {\bibfnamefont {J.}~\bibnamefont {Zhu}}, \bibinfo {author} {\bibfnamefont {W.}~\bibnamefont {Wu}}, \bibinfo {author} {\bibfnamefont {C.}~\bibnamefont {Xiao}},\ and\ \bibinfo {author} {\bibfnamefont {S.~A.}\ \bibnamefont {Yang}},\ }\href {https://doi.org/10.1103/PhysRevLett.132.056301} {\bibfield  {journal} {\bibinfo  {journal} {Phys. Rev. Lett.}\ }\textbf {\bibinfo {volume} {132}},\ \bibinfo {pages} {056301} (\bibinfo {year} {2024})}\BibitemShut {NoStop}%
\bibitem [{\citenamefont {Wang}\ \emph {et~al.}(2013)\citenamefont {Wang}, \citenamefont {Weng}, \citenamefont {Wu}, \citenamefont {Dai},\ and\ \citenamefont {Fang}}]{wang_three-dimensional_2013}%
  \BibitemOpen
  \bibfield  {author} {\bibinfo {author} {\bibfnamefont {Z.}~\bibnamefont {Wang}}, \bibinfo {author} {\bibfnamefont {H.}~\bibnamefont {Weng}}, \bibinfo {author} {\bibfnamefont {Q.}~\bibnamefont {Wu}}, \bibinfo {author} {\bibfnamefont {X.}~\bibnamefont {Dai}},\ and\ \bibinfo {author} {\bibfnamefont {Z.}~\bibnamefont {Fang}},\ }\href {https://doi.org/10.1103/PhysRevB.88.125427} {\bibfield  {journal} {\bibinfo  {journal} {Phys. Rev. B}\ }\textbf {\bibinfo {volume} {88}},\ \bibinfo {pages} {125427} (\bibinfo {year} {2013})}\BibitemShut {NoStop}%
\bibitem [{\citenamefont {Ali}\ \emph {et~al.}(2014)\citenamefont {Ali}, \citenamefont {Gibson}, \citenamefont {Jeon}, \citenamefont {Zhou}, \citenamefont {Yazdani},\ and\ \citenamefont {Cava}}]{ali_crystal_2014}%
  \BibitemOpen
  \bibfield  {author} {\bibinfo {author} {\bibfnamefont {M.~N.}\ \bibnamefont {Ali}}, \bibinfo {author} {\bibfnamefont {Q.}~\bibnamefont {Gibson}}, \bibinfo {author} {\bibfnamefont {S.}~\bibnamefont {Jeon}}, \bibinfo {author} {\bibfnamefont {B.~B.}\ \bibnamefont {Zhou}}, \bibinfo {author} {\bibfnamefont {A.}~\bibnamefont {Yazdani}},\ and\ \bibinfo {author} {\bibfnamefont {R.~J.}\ \bibnamefont {Cava}},\ }\href {https://doi.org/10.1021/ic403163d} {\bibfield  {journal} {\bibinfo  {journal} {Inorg. Chem.}\ }\textbf {\bibinfo {volume} {53}},\ \bibinfo {pages} {4062} (\bibinfo {year} {2014})}\BibitemShut {NoStop}%
\bibitem [{\citenamefont {Sankar}\ \emph {et~al.}(2015)\citenamefont {Sankar}, \citenamefont {Neupane}, \citenamefont {Xu}, \citenamefont {Butler}, \citenamefont {Zeljkovic}, \citenamefont {Panneer~Muthuselvam}, \citenamefont {Huang}, \citenamefont {Guo}, \citenamefont {Karna}, \citenamefont {Chu}, \citenamefont {Lee}, \citenamefont {Lin}, \citenamefont {Jayavel}, \citenamefont {Madhavan}, \citenamefont {Hasan},\ and\ \citenamefont {Chou}}]{Sankar2015-je}%
  \BibitemOpen
  \bibfield  {author} {\bibinfo {author} {\bibfnamefont {R.}~\bibnamefont {Sankar}}, \bibinfo {author} {\bibfnamefont {M.}~\bibnamefont {Neupane}}, \bibinfo {author} {\bibfnamefont {S.-Y.}\ \bibnamefont {Xu}}, \bibinfo {author} {\bibfnamefont {C.~J.}\ \bibnamefont {Butler}}, \bibinfo {author} {\bibfnamefont {I.}~\bibnamefont {Zeljkovic}}, \bibinfo {author} {\bibfnamefont {I.}~\bibnamefont {Panneer~Muthuselvam}}, \bibinfo {author} {\bibfnamefont {F.-T.}\ \bibnamefont {Huang}}, \bibinfo {author} {\bibfnamefont {S.-T.}\ \bibnamefont {Guo}}, \bibinfo {author} {\bibfnamefont {S.~K.}\ \bibnamefont {Karna}}, \bibinfo {author} {\bibfnamefont {M.-W.}\ \bibnamefont {Chu}}, \bibinfo {author} {\bibfnamefont {W.~L.}\ \bibnamefont {Lee}}, \bibinfo {author} {\bibfnamefont {M.-T.}\ \bibnamefont {Lin}}, \bibinfo {author} {\bibfnamefont {R.}~\bibnamefont {Jayavel}}, \bibinfo {author} {\bibfnamefont {V.}~\bibnamefont {Madhavan}}, \bibinfo {author} {\bibfnamefont {M.~Z.}\ \bibnamefont {Hasan}},\ and\ \bibinfo {author}
  {\bibfnamefont {F.~C.}\ \bibnamefont {Chou}},\ }\href@noop {} {\bibfield  {journal} {\bibinfo  {journal} {Sci. Rep.}\ }\textbf {\bibinfo {volume} {5}},\ \bibinfo {pages} {12966} (\bibinfo {year} {2015})}\BibitemShut {NoStop}%
\bibitem [{\citenamefont {Liu}\ \emph {et~al.}(2010{\natexlab{a}})\citenamefont {Liu}, \citenamefont {Qi}, \citenamefont {Zhang}, \citenamefont {Dai}, \citenamefont {Fang},\ and\ \citenamefont {Zhang}}]{liu_model_2010}%
  \BibitemOpen
  \bibfield  {author} {\bibinfo {author} {\bibfnamefont {C.-X.}\ \bibnamefont {Liu}}, \bibinfo {author} {\bibfnamefont {X.-L.}\ \bibnamefont {Qi}}, \bibinfo {author} {\bibfnamefont {H.}~\bibnamefont {Zhang}}, \bibinfo {author} {\bibfnamefont {X.}~\bibnamefont {Dai}}, \bibinfo {author} {\bibfnamefont {Z.}~\bibnamefont {Fang}},\ and\ \bibinfo {author} {\bibfnamefont {S.-C.}\ \bibnamefont {Zhang}},\ }\href {https://doi.org/10.1103/PhysRevB.82.045122} {\bibfield  {journal} {\bibinfo  {journal} {Phys. Rev. B}\ }\textbf {\bibinfo {volume} {82}},\ \bibinfo {pages} {045122} (\bibinfo {year} {2010}{\natexlab{a}})}\BibitemShut {NoStop}%
\bibitem [{\citenamefont {Liu}\ \emph {et~al.}(2010{\natexlab{b}})\citenamefont {Liu}, \citenamefont {Zhang}, \citenamefont {Yan}, \citenamefont {Qi}, \citenamefont {Frauenheim}, \citenamefont {Dai}, \citenamefont {Fang},\ and\ \citenamefont {Zhang}}]{chaoxing_2DTI_crossover}%
  \BibitemOpen
  \bibfield  {author} {\bibinfo {author} {\bibfnamefont {C.-X.}\ \bibnamefont {Liu}}, \bibinfo {author} {\bibfnamefont {H.}~\bibnamefont {Zhang}}, \bibinfo {author} {\bibfnamefont {B.}~\bibnamefont {Yan}}, \bibinfo {author} {\bibfnamefont {X.-L.}\ \bibnamefont {Qi}}, \bibinfo {author} {\bibfnamefont {T.}~\bibnamefont {Frauenheim}}, \bibinfo {author} {\bibfnamefont {X.}~\bibnamefont {Dai}}, \bibinfo {author} {\bibfnamefont {Z.}~\bibnamefont {Fang}},\ and\ \bibinfo {author} {\bibfnamefont {S.-C.}\ \bibnamefont {Zhang}},\ }\href {https://doi.org/10.1103/PhysRevB.81.041307} {\bibfield  {journal} {\bibinfo  {journal} {Phys. Rev. B}\ }\textbf {\bibinfo {volume} {81}},\ \bibinfo {pages} {041307} (\bibinfo {year} {2010}{\natexlab{b}})}\BibitemShut {NoStop}%
\bibitem [{\citenamefont {Lu}\ \emph {et~al.}(2010)\citenamefont {Lu}, \citenamefont {Shan}, \citenamefont {Yao}, \citenamefont {Niu},\ and\ \citenamefont {Shen}}]{lu_ti_2d}%
  \BibitemOpen
  \bibfield  {author} {\bibinfo {author} {\bibfnamefont {H.-Z.}\ \bibnamefont {Lu}}, \bibinfo {author} {\bibfnamefont {W.-Y.}\ \bibnamefont {Shan}}, \bibinfo {author} {\bibfnamefont {W.}~\bibnamefont {Yao}}, \bibinfo {author} {\bibfnamefont {Q.}~\bibnamefont {Niu}},\ and\ \bibinfo {author} {\bibfnamefont {S.-Q.}\ \bibnamefont {Shen}},\ }\href {https://doi.org/10.1103/PhysRevB.81.115407} {\bibfield  {journal} {\bibinfo  {journal} {Phys. Rev. B}\ }\textbf {\bibinfo {volume} {81}},\ \bibinfo {pages} {115407} (\bibinfo {year} {2010})}\BibitemShut {NoStop}%
\bibitem [{\citenamefont {Linder}\ \emph {et~al.}(2009)\citenamefont {Linder}, \citenamefont {Yokoyama},\ and\ \citenamefont {Sudb\o{}}}]{finite_size}%
  \BibitemOpen
  \bibfield  {author} {\bibinfo {author} {\bibfnamefont {J.}~\bibnamefont {Linder}}, \bibinfo {author} {\bibfnamefont {T.}~\bibnamefont {Yokoyama}},\ and\ \bibinfo {author} {\bibfnamefont {A.}~\bibnamefont {Sudb\o{}}},\ }\href {https://doi.org/10.1103/PhysRevB.80.205401} {\bibfield  {journal} {\bibinfo  {journal} {Phys. Rev. B}\ }\textbf {\bibinfo {volume} {80}},\ \bibinfo {pages} {205401} (\bibinfo {year} {2009})}\BibitemShut {NoStop}%
\bibitem [{\citenamefont {Chang}\ \emph {et~al.}(2013)\citenamefont {Chang}, \citenamefont {Zhang}, \citenamefont {Feng}, \citenamefont {Shen}, \citenamefont {Zhang}, \citenamefont {Guo}, \citenamefont {Li}, \citenamefont {Ou}, \citenamefont {Wei}, \citenamefont {Wang} \emph {et~al.}}]{chang2013experimental}%
  \BibitemOpen
  \bibfield  {author} {\bibinfo {author} {\bibfnamefont {C.-Z.}\ \bibnamefont {Chang}}, \bibinfo {author} {\bibfnamefont {J.}~\bibnamefont {Zhang}}, \bibinfo {author} {\bibfnamefont {X.}~\bibnamefont {Feng}}, \bibinfo {author} {\bibfnamefont {J.}~\bibnamefont {Shen}}, \bibinfo {author} {\bibfnamefont {Z.}~\bibnamefont {Zhang}}, \bibinfo {author} {\bibfnamefont {M.}~\bibnamefont {Guo}}, \bibinfo {author} {\bibfnamefont {K.}~\bibnamefont {Li}}, \bibinfo {author} {\bibfnamefont {Y.}~\bibnamefont {Ou}}, \bibinfo {author} {\bibfnamefont {P.}~\bibnamefont {Wei}}, \bibinfo {author} {\bibfnamefont {L.-L.}\ \bibnamefont {Wang}}, \emph {et~al.},\ }\href@noop {} {\bibfield  {journal} {\bibinfo  {journal} {Science}\ }\textbf {\bibinfo {volume} {340}},\ \bibinfo {pages} {167} (\bibinfo {year} {2013})}\BibitemShut {NoStop}%
\bibitem [{\citenamefont {Deng}\ \emph {et~al.}(2020)\citenamefont {Deng}, \citenamefont {Yu}, \citenamefont {Shi}, \citenamefont {Guo}, \citenamefont {Xu}, \citenamefont {Wang}, \citenamefont {Chen},\ and\ \citenamefont {Zhang}}]{deng2020quantum}%
  \BibitemOpen
  \bibfield  {author} {\bibinfo {author} {\bibfnamefont {Y.}~\bibnamefont {Deng}}, \bibinfo {author} {\bibfnamefont {Y.}~\bibnamefont {Yu}}, \bibinfo {author} {\bibfnamefont {M.~Z.}\ \bibnamefont {Shi}}, \bibinfo {author} {\bibfnamefont {Z.}~\bibnamefont {Guo}}, \bibinfo {author} {\bibfnamefont {Z.}~\bibnamefont {Xu}}, \bibinfo {author} {\bibfnamefont {J.}~\bibnamefont {Wang}}, \bibinfo {author} {\bibfnamefont {X.~H.}\ \bibnamefont {Chen}},\ and\ \bibinfo {author} {\bibfnamefont {Y.}~\bibnamefont {Zhang}},\ }\href@noop {} {\bibfield  {journal} {\bibinfo  {journal} {Science}\ }\textbf {\bibinfo {volume} {367}},\ \bibinfo {pages} {895} (\bibinfo {year} {2020})}\BibitemShut {NoStop}%
\bibitem [{\citenamefont {Sharpe}\ \emph {et~al.}(2019)\citenamefont {Sharpe}, \citenamefont {Fox}, \citenamefont {Barnard}, \citenamefont {Finney}, \citenamefont {Watanabe}, \citenamefont {Taniguchi}, \citenamefont {Kastner},\ and\ \citenamefont {Goldhaber-Gordon}}]{sharpe2019emergent}%
  \BibitemOpen
  \bibfield  {author} {\bibinfo {author} {\bibfnamefont {A.~L.}\ \bibnamefont {Sharpe}}, \bibinfo {author} {\bibfnamefont {E.~J.}\ \bibnamefont {Fox}}, \bibinfo {author} {\bibfnamefont {A.~W.}\ \bibnamefont {Barnard}}, \bibinfo {author} {\bibfnamefont {J.}~\bibnamefont {Finney}}, \bibinfo {author} {\bibfnamefont {K.}~\bibnamefont {Watanabe}}, \bibinfo {author} {\bibfnamefont {T.}~\bibnamefont {Taniguchi}}, \bibinfo {author} {\bibfnamefont {M.}~\bibnamefont {Kastner}},\ and\ \bibinfo {author} {\bibfnamefont {D.}~\bibnamefont {Goldhaber-Gordon}},\ }\href@noop {} {\bibfield  {journal} {\bibinfo  {journal} {Science}\ }\textbf {\bibinfo {volume} {365}},\ \bibinfo {pages} {605} (\bibinfo {year} {2019})}\BibitemShut {NoStop}%
\bibitem [{\citenamefont {Serlin}\ \emph {et~al.}(2020)\citenamefont {Serlin}, \citenamefont {Tschirhart}, \citenamefont {Polshyn}, \citenamefont {Zhang}, \citenamefont {Zhu}, \citenamefont {Watanabe}, \citenamefont {Taniguchi}, \citenamefont {Balents},\ and\ \citenamefont {Young}}]{serlin2020intrinsic}%
  \BibitemOpen
  \bibfield  {author} {\bibinfo {author} {\bibfnamefont {M.}~\bibnamefont {Serlin}}, \bibinfo {author} {\bibfnamefont {C.}~\bibnamefont {Tschirhart}}, \bibinfo {author} {\bibfnamefont {H.}~\bibnamefont {Polshyn}}, \bibinfo {author} {\bibfnamefont {Y.}~\bibnamefont {Zhang}}, \bibinfo {author} {\bibfnamefont {J.}~\bibnamefont {Zhu}}, \bibinfo {author} {\bibfnamefont {K.}~\bibnamefont {Watanabe}}, \bibinfo {author} {\bibfnamefont {T.}~\bibnamefont {Taniguchi}}, \bibinfo {author} {\bibfnamefont {L.}~\bibnamefont {Balents}},\ and\ \bibinfo {author} {\bibfnamefont {A.}~\bibnamefont {Young}},\ }\href@noop {} {\bibfield  {journal} {\bibinfo  {journal} {Science}\ }\textbf {\bibinfo {volume} {367}},\ \bibinfo {pages} {900} (\bibinfo {year} {2020})}\BibitemShut {NoStop}%
\bibitem [{\citenamefont {Chen}\ \emph {et~al.}(2020)\citenamefont {Chen}, \citenamefont {Sharpe}, \citenamefont {Fox}, \citenamefont {Zhang}, \citenamefont {Wang}, \citenamefont {Jiang}, \citenamefont {Lyu}, \citenamefont {Li}, \citenamefont {Watanabe}, \citenamefont {Taniguchi} \emph {et~al.}}]{chen2020tunable}%
  \BibitemOpen
  \bibfield  {author} {\bibinfo {author} {\bibfnamefont {G.}~\bibnamefont {Chen}}, \bibinfo {author} {\bibfnamefont {A.~L.}\ \bibnamefont {Sharpe}}, \bibinfo {author} {\bibfnamefont {E.~J.}\ \bibnamefont {Fox}}, \bibinfo {author} {\bibfnamefont {Y.-H.}\ \bibnamefont {Zhang}}, \bibinfo {author} {\bibfnamefont {S.}~\bibnamefont {Wang}}, \bibinfo {author} {\bibfnamefont {L.}~\bibnamefont {Jiang}}, \bibinfo {author} {\bibfnamefont {B.}~\bibnamefont {Lyu}}, \bibinfo {author} {\bibfnamefont {H.}~\bibnamefont {Li}}, \bibinfo {author} {\bibfnamefont {K.}~\bibnamefont {Watanabe}}, \bibinfo {author} {\bibfnamefont {T.}~\bibnamefont {Taniguchi}}, \emph {et~al.},\ }\href@noop {} {\bibfield  {journal} {\bibinfo  {journal} {Nature}\ }\textbf {\bibinfo {volume} {579}},\ \bibinfo {pages} {56} (\bibinfo {year} {2020})}\BibitemShut {NoStop}%
\bibitem [{\citenamefont {Li}\ \emph {et~al.}(2021)\citenamefont {Li}, \citenamefont {Jiang}, \citenamefont {Shen}, \citenamefont {Zhang}, \citenamefont {Li}, \citenamefont {Tao}, \citenamefont {Devakul}, \citenamefont {Watanabe}, \citenamefont {Taniguchi}, \citenamefont {Fu} \emph {et~al.}}]{li2021quantum}%
  \BibitemOpen
  \bibfield  {author} {\bibinfo {author} {\bibfnamefont {T.}~\bibnamefont {Li}}, \bibinfo {author} {\bibfnamefont {S.}~\bibnamefont {Jiang}}, \bibinfo {author} {\bibfnamefont {B.}~\bibnamefont {Shen}}, \bibinfo {author} {\bibfnamefont {Y.}~\bibnamefont {Zhang}}, \bibinfo {author} {\bibfnamefont {L.}~\bibnamefont {Li}}, \bibinfo {author} {\bibfnamefont {Z.}~\bibnamefont {Tao}}, \bibinfo {author} {\bibfnamefont {T.}~\bibnamefont {Devakul}}, \bibinfo {author} {\bibfnamefont {K.}~\bibnamefont {Watanabe}}, \bibinfo {author} {\bibfnamefont {T.}~\bibnamefont {Taniguchi}}, \bibinfo {author} {\bibfnamefont {L.}~\bibnamefont {Fu}}, \emph {et~al.},\ }\href@noop {} {\bibfield  {journal} {\bibinfo  {journal} {Nature}\ }\textbf {\bibinfo {volume} {600}},\ \bibinfo {pages} {641} (\bibinfo {year} {2021})}\BibitemShut {NoStop}%
\bibitem [{\citenamefont {Chang}\ \emph {et~al.}(2023)\citenamefont {Chang}, \citenamefont {Liu},\ and\ \citenamefont {MacDonald}}]{chang_liu_rmp}%
  \BibitemOpen
  \bibfield  {author} {\bibinfo {author} {\bibfnamefont {C.-Z.}\ \bibnamefont {Chang}}, \bibinfo {author} {\bibfnamefont {C.-X.}\ \bibnamefont {Liu}},\ and\ \bibinfo {author} {\bibfnamefont {A.~H.}\ \bibnamefont {MacDonald}},\ }\href {https://doi.org/10.1103/RevModPhys.95.011002} {\bibfield  {journal} {\bibinfo  {journal} {Rev. Mod. Phys.}\ }\textbf {\bibinfo {volume} {95}},\ \bibinfo {pages} {011002} (\bibinfo {year} {2023})}\BibitemShut {NoStop}%
\bibitem [{\citenamefont {Fu}(2009)}]{liang_hex}%
  \BibitemOpen
  \bibfield  {author} {\bibinfo {author} {\bibfnamefont {L.}~\bibnamefont {Fu}},\ }\href {https://doi.org/10.1103/PhysRevLett.103.266801} {\bibfield  {journal} {\bibinfo  {journal} {Phys. Rev. Lett.}\ }\textbf {\bibinfo {volume} {103}},\ \bibinfo {pages} {266801} (\bibinfo {year} {2009})}\BibitemShut {NoStop}%
\bibitem [{\citenamefont {Wang}\ \emph {et~al.}(2021)\citenamefont {Wang}, \citenamefont {Yuan},\ and\ \citenamefont {Fu}}]{moire_surf_fu}%
  \BibitemOpen
  \bibfield  {author} {\bibinfo {author} {\bibfnamefont {T.}~\bibnamefont {Wang}}, \bibinfo {author} {\bibfnamefont {N.~F.~Q.}\ \bibnamefont {Yuan}},\ and\ \bibinfo {author} {\bibfnamefont {L.}~\bibnamefont {Fu}},\ }\href {https://doi.org/10.1103/PhysRevX.11.021024} {\bibfield  {journal} {\bibinfo  {journal} {Phys. Rev. X}\ }\textbf {\bibinfo {volume} {11}},\ \bibinfo {pages} {021024} (\bibinfo {year} {2021})}\BibitemShut {NoStop}%
\bibitem [{\citenamefont {Cano}\ \emph {et~al.}(2021)\citenamefont {Cano}, \citenamefont {Fang}, \citenamefont {Pixley},\ and\ \citenamefont {Wilson}}]{cano_moire_surface}%
  \BibitemOpen
  \bibfield  {author} {\bibinfo {author} {\bibfnamefont {J.}~\bibnamefont {Cano}}, \bibinfo {author} {\bibfnamefont {S.}~\bibnamefont {Fang}}, \bibinfo {author} {\bibfnamefont {J.~H.}\ \bibnamefont {Pixley}},\ and\ \bibinfo {author} {\bibfnamefont {J.~H.}\ \bibnamefont {Wilson}},\ }\href {https://doi.org/10.1103/PhysRevB.103.155157} {\bibfield  {journal} {\bibinfo  {journal} {Phys. Rev. B}\ }\textbf {\bibinfo {volume} {103}},\ \bibinfo {pages} {155157} (\bibinfo {year} {2021})}\BibitemShut {NoStop}%
\bibitem [{\citenamefont {Ghorashi}\ \emph {et~al.}(2023)\citenamefont {Ghorashi}, \citenamefont {Dunbrack}, \citenamefont {Abouelkomsan}, \citenamefont {Sun}, \citenamefont {Du},\ and\ \citenamefont {Cano}}]{moire_pont_cano}%
  \BibitemOpen
  \bibfield  {author} {\bibinfo {author} {\bibfnamefont {S.~A.~A.}\ \bibnamefont {Ghorashi}}, \bibinfo {author} {\bibfnamefont {A.}~\bibnamefont {Dunbrack}}, \bibinfo {author} {\bibfnamefont {A.}~\bibnamefont {Abouelkomsan}}, \bibinfo {author} {\bibfnamefont {J.}~\bibnamefont {Sun}}, \bibinfo {author} {\bibfnamefont {X.}~\bibnamefont {Du}},\ and\ \bibinfo {author} {\bibfnamefont {J.}~\bibnamefont {Cano}},\ }\href {https://doi.org/10.1103/PhysRevLett.130.196201} {\bibfield  {journal} {\bibinfo  {journal} {Phys. Rev. Lett.}\ }\textbf {\bibinfo {volume} {130}},\ \bibinfo {pages} {196201} (\bibinfo {year} {2023})}\BibitemShut {NoStop}%
\bibitem [{\citenamefont {Wan}\ \emph {et~al.}(2023)\citenamefont {Wan}, \citenamefont {Sarkar}, \citenamefont {Lin},\ and\ \citenamefont {Sun}}]{wan_moire_strain}%
  \BibitemOpen
  \bibfield  {author} {\bibinfo {author} {\bibfnamefont {X.}~\bibnamefont {Wan}}, \bibinfo {author} {\bibfnamefont {S.}~\bibnamefont {Sarkar}}, \bibinfo {author} {\bibfnamefont {S.-Z.}\ \bibnamefont {Lin}},\ and\ \bibinfo {author} {\bibfnamefont {K.}~\bibnamefont {Sun}},\ }\href {https://doi.org/10.1103/PhysRevLett.130.216401} {\bibfield  {journal} {\bibinfo  {journal} {Phys. Rev. Lett.}\ }\textbf {\bibinfo {volume} {130}},\ \bibinfo {pages} {216401} (\bibinfo {year} {2023})}\BibitemShut {NoStop}%
\bibitem [{\citenamefont {Su}\ \emph {et~al.}(2022)\citenamefont {Su}, \citenamefont {Li}, \citenamefont {Zhang}, \citenamefont {Sun},\ and\ \citenamefont {Lin}}]{moire-dirac_lin_prr_2023}%
  \BibitemOpen
  \bibfield  {author} {\bibinfo {author} {\bibfnamefont {Y.}~\bibnamefont {Su}}, \bibinfo {author} {\bibfnamefont {H.}~\bibnamefont {Li}}, \bibinfo {author} {\bibfnamefont {C.}~\bibnamefont {Zhang}}, \bibinfo {author} {\bibfnamefont {K.}~\bibnamefont {Sun}},\ and\ \bibinfo {author} {\bibfnamefont {S.-Z.}\ \bibnamefont {Lin}},\ }\href {https://doi.org/10.1103/PhysRevResearch.4.L032024} {\bibfield  {journal} {\bibinfo  {journal} {Phys. Rev. Res.}\ }\textbf {\bibinfo {volume} {4}},\ \bibinfo {pages} {L032024} (\bibinfo {year} {2022})}\BibitemShut {NoStop}%
\bibitem [{\citenamefont {Miao}\ \emph {et~al.}(2023)\citenamefont {Miao}, \citenamefont {Li}, \citenamefont {Han}, \citenamefont {Pan},\ and\ \citenamefont {Dai}}]{miao_truncated}%
  \BibitemOpen
  \bibfield  {author} {\bibinfo {author} {\bibfnamefont {W.}~\bibnamefont {Miao}}, \bibinfo {author} {\bibfnamefont {C.}~\bibnamefont {Li}}, \bibinfo {author} {\bibfnamefont {X.}~\bibnamefont {Han}}, \bibinfo {author} {\bibfnamefont {D.}~\bibnamefont {Pan}},\ and\ \bibinfo {author} {\bibfnamefont {X.}~\bibnamefont {Dai}},\ }\href {https://doi.org/10.1103/PhysRevB.107.125112} {\bibfield  {journal} {\bibinfo  {journal} {Phys. Rev. B}\ }\textbf {\bibinfo {volume} {107}},\ \bibinfo {pages} {125112} (\bibinfo {year} {2023})}\BibitemShut {NoStop}%
\bibitem [{\citenamefont {Sun}\ \emph {et~al.}(2023)\citenamefont {Sun}, \citenamefont {Ghorashi}, \citenamefont {Watanabe}, \citenamefont {Taniguchi}, \citenamefont {Camino}, \citenamefont {Cano},\ and\ \citenamefont {Du}}]{sun2023signature}%
  \BibitemOpen
  \bibfield  {author} {\bibinfo {author} {\bibfnamefont {J.}~\bibnamefont {Sun}}, \bibinfo {author} {\bibfnamefont {S.~A.~A.}\ \bibnamefont {Ghorashi}}, \bibinfo {author} {\bibfnamefont {K.}~\bibnamefont {Watanabe}}, \bibinfo {author} {\bibfnamefont {T.}~\bibnamefont {Taniguchi}}, \bibinfo {author} {\bibfnamefont {F.}~\bibnamefont {Camino}}, \bibinfo {author} {\bibfnamefont {J.}~\bibnamefont {Cano}},\ and\ \bibinfo {author} {\bibfnamefont {X.}~\bibnamefont {Du}},\ }\href@noop {} {\bibfield  {journal} {\bibinfo  {journal} {arXiv preprint arXiv:2306.06848}\ } (\bibinfo {year} {2023})}\BibitemShut {NoStop}%
\bibitem [{\citenamefont {Kim}\ \emph {et~al.}(2023)\citenamefont {Kim}, \citenamefont {Dominguez}, \citenamefont {Mayorga-Luna}, \citenamefont {Ye}, \citenamefont {Embley}, \citenamefont {Tan}, \citenamefont {Ni}, \citenamefont {Liu}, \citenamefont {Ford}, \citenamefont {Gao}, \citenamefont {Arash}, \citenamefont {Watanabe}, \citenamefont {Taniguchi}, \citenamefont {Kim}, \citenamefont {Shih}, \citenamefont {Lai}, \citenamefont {Yao}, \citenamefont {Yang}, \citenamefont {Li},\ and\ \citenamefont {Miyahara}}]{Kim2023_thbn}%
  \BibitemOpen
  \bibfield  {author} {\bibinfo {author} {\bibfnamefont {D.~S.}\ \bibnamefont {Kim}}, \bibinfo {author} {\bibfnamefont {R.~C.}\ \bibnamefont {Dominguez}}, \bibinfo {author} {\bibfnamefont {R.}~\bibnamefont {Mayorga-Luna}}, \bibinfo {author} {\bibfnamefont {D.}~\bibnamefont {Ye}}, \bibinfo {author} {\bibfnamefont {J.}~\bibnamefont {Embley}}, \bibinfo {author} {\bibfnamefont {T.}~\bibnamefont {Tan}}, \bibinfo {author} {\bibfnamefont {Y.}~\bibnamefont {Ni}}, \bibinfo {author} {\bibfnamefont {Z.}~\bibnamefont {Liu}}, \bibinfo {author} {\bibfnamefont {M.}~\bibnamefont {Ford}}, \bibinfo {author} {\bibfnamefont {F.~Y.}\ \bibnamefont {Gao}}, \bibinfo {author} {\bibfnamefont {S.}~\bibnamefont {Arash}}, \bibinfo {author} {\bibfnamefont {K.}~\bibnamefont {Watanabe}}, \bibinfo {author} {\bibfnamefont {T.}~\bibnamefont {Taniguchi}}, \bibinfo {author} {\bibfnamefont {S.}~\bibnamefont {Kim}}, \bibinfo {author} {\bibfnamefont {C.-K.}\ \bibnamefont {Shih}}, \bibinfo {author} {\bibfnamefont {K.}~\bibnamefont {Lai}}, \bibinfo
  {author} {\bibfnamefont {W.}~\bibnamefont {Yao}}, \bibinfo {author} {\bibfnamefont {L.}~\bibnamefont {Yang}}, \bibinfo {author} {\bibfnamefont {X.}~\bibnamefont {Li}},\ and\ \bibinfo {author} {\bibfnamefont {Y.}~\bibnamefont {Miyahara}},\ }\bibfield  {journal} {\bibinfo  {journal} {Nature Materials}\ }\href {https://doi.org/10.1038/s41563-023-01637-7} {10.1038/s41563-023-01637-7} (\bibinfo {year} {2023})\BibitemShut {NoStop}%
\bibitem [{\citenamefont {Battilomo}\ \emph {et~al.}(2021)\citenamefont {Battilomo}, \citenamefont {Scopigno},\ and\ \citenamefont {Ortix}}]{anomalous_planar_hall_prr}%
  \BibitemOpen
  \bibfield  {author} {\bibinfo {author} {\bibfnamefont {R.}~\bibnamefont {Battilomo}}, \bibinfo {author} {\bibfnamefont {N.}~\bibnamefont {Scopigno}},\ and\ \bibinfo {author} {\bibfnamefont {C.}~\bibnamefont {Ortix}},\ }\href {https://doi.org/10.1103/PhysRevResearch.3.L012006} {\bibfield  {journal} {\bibinfo  {journal} {Phys. Rev. Res.}\ }\textbf {\bibinfo {volume} {3}},\ \bibinfo {pages} {L012006} (\bibinfo {year} {2021})}\BibitemShut {NoStop}%
\bibitem [{\citenamefont {Wang}\ \emph {et~al.}(2023)\citenamefont {Wang}, \citenamefont {Du}, \citenamefont {Lu},\ and\ \citenamefont {Xie}}]{absence_aphe}%
  \BibitemOpen
  \bibfield  {author} {\bibinfo {author} {\bibfnamefont {C.~M.}\ \bibnamefont {Wang}}, \bibinfo {author} {\bibfnamefont {Z.~Z.}\ \bibnamefont {Du}}, \bibinfo {author} {\bibfnamefont {H.-Z.}\ \bibnamefont {Lu}},\ and\ \bibinfo {author} {\bibfnamefont {X.~C.}\ \bibnamefont {Xie}},\ }\href {https://doi.org/10.1103/PhysRevB.108.L121301} {\bibfield  {journal} {\bibinfo  {journal} {Phys. Rev. B}\ }\textbf {\bibinfo {volume} {108}},\ \bibinfo {pages} {L121301} (\bibinfo {year} {2023})}\BibitemShut {NoStop}%
\bibitem [{\citenamefont {Huang}\ \emph {et~al.}(2023{\natexlab{b}})\citenamefont {Huang}, \citenamefont {Feng}, \citenamefont {Wang}, \citenamefont {Xiao},\ and\ \citenamefont {Yang}}]{intrinsic_nonlinear_planar_hall_prl}%
  \BibitemOpen
  \bibfield  {author} {\bibinfo {author} {\bibfnamefont {Y.-X.}\ \bibnamefont {Huang}}, \bibinfo {author} {\bibfnamefont {X.}~\bibnamefont {Feng}}, \bibinfo {author} {\bibfnamefont {H.}~\bibnamefont {Wang}}, \bibinfo {author} {\bibfnamefont {C.}~\bibnamefont {Xiao}},\ and\ \bibinfo {author} {\bibfnamefont {S.~A.}\ \bibnamefont {Yang}},\ }\href {https://doi.org/10.1103/PhysRevLett.130.126303} {\bibfield  {journal} {\bibinfo  {journal} {Phys. Rev. Lett.}\ }\textbf {\bibinfo {volume} {130}},\ \bibinfo {pages} {126303} (\bibinfo {year} {2023}{\natexlab{b}})}\BibitemShut {NoStop}%
\end{thebibliography}%

\end{document}